\documentclass[journal,compsoc]{IEEEtran}                   
\usepackage{orcidlink}
\graphicspath{{figs/}{figures/}{pictures/}{images/}{./}} 

\usepackage{subcaption}
\usepackage{booktabs}                  
\usepackage{amsmath,amssymb}           
\usepackage{mathptmx}                  
\usepackage{nicefrac}
\usepackage{multirow}
\usepackage{adjustbox}
\newcommand{\etal}{\textit{et al.}}

\usepackage{xcolor}

\long\def\highlight#1{%
  {\color{black}#1}%
}

\begin{document}
\title{VISTA: Differentiable Viewpoint Planning for Long-Horizon Task Visualization}
\author{Donggang Jia\orcidlink{0000-0002-1358-8718}, Da Li\orcidlink{0009-0009-9914-2503}, Zainab Alsuwaykit\orcidlink{0009-0009-7444-445X}, Sebastian Hartwig\orcidlink{0000-0001-8642-2789}, \\Dominik Engel\orcidlink{0000-0002-5766-7215}, Timo Ropinski\orcidlink{0000-0002-7857-5512} and Ivan Viola\orcidlink{0000-0003-4248-6574}

\thanks{Donggang Jia, Da Li, Zainab Alsuwaykit, Dominik Engel and Ivan Viola are with the Computer, Electrical and Mathematical Sciences and Engineering Division, King Abdullah University of Science and Technology, Saudi Arabia. E-mail:\{donggang.jia\,$|$\,da.li\,$|$\,zainab.alsuwaykit\,$|$\,dominik.engel\,$|$\,\\ivan.viola\}@kaust.edu.sa.}
\thanks{Sebastian Hartwig is with the Liebherr Digital Development Center GmbH, Germany. E-mail: sebastian@hartwig-bw.de.}
\thanks{Timo Ropinski is with the Visual Computing Group, Ulm University, Germany. E-mail: timo.ropinski@uni-ulm.de.}
\thanks{Manuscript received April 19, 2021; revised August 16, 2021.}}

\markboth{Journal of \LaTeX\ Class Files,~Vol.~14, No.~8, August~2021}%
{Shell \MakeLowercase{\textit{et al.}}: A Sample Article Using IEEEtran.cls for IEEE Journals}

\IEEEpubid{0000--0000/00\$00.00~\copyright~2021 IEEE}

\maketitle
\begin{abstract}
\highlight{Long-horizon object-centered tasks in interactive 3D environments involve heterogeneous events that unfold across time and space, including fine-grained manipulations, larger structural motions, and subtle state transitions.} Effectively visualizing such processes is essential for helping users understand how actions lead to outcomes over time. However, existing camera control methods in 3D systems are commonly designed for static objects, and often fail to maintain the legibility of task-relevant components and their evolving relationships throughout long-horizon executions. We present VISTA, a differentiable viewpoint planning framework for long-horizon task visualization. We formulate camera control as a visualization policy that optimizes camera parameters under a unified rendering-based quality objective and we propose five quality metrics to satisfy our five design goals. To support both local part inspection and interaction-centric visualization centered on a primary object, our formulation handles targets defined either as object--part pairs or as relational tuples involving another interacting object. This allows the planner to reveal not only what part is important, but also how it functionally relates to surrounding structures during task execution. We demonstrate that the same planner supports two complementary use cases: presentation-driven narrative visualization for communicating how long-horizon tasks unfold to human viewers, and agent-driven active observation within an observe-act loop. \highlight{Together, these results position differentiable viewpoint planning as a general mechanism for ``how-things-work'' visualization of long-horizon object-centered interactions in 3D environments.}
\end{abstract}

\section{Introduction} \label{sec:introduction} 
\maketitle
\begin{figure*}
    \centering
    \includegraphics[width=\linewidth]{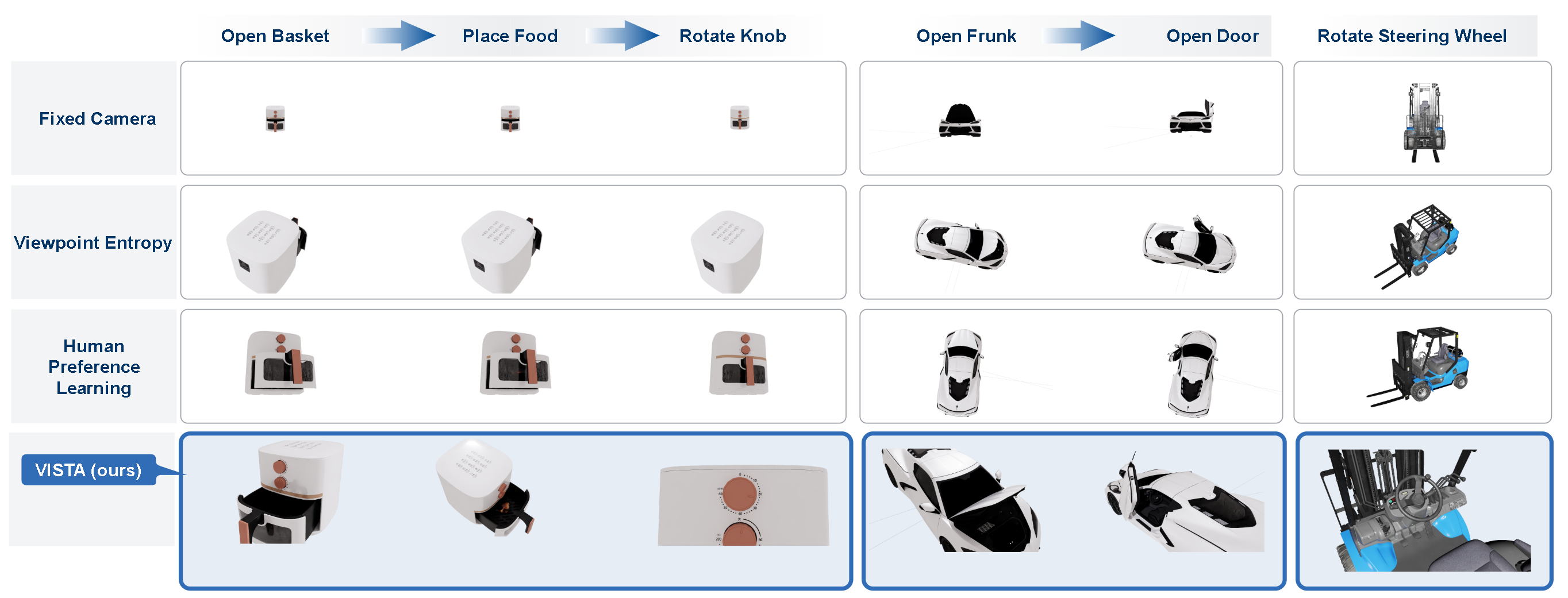}
    \caption{\textbf{Differentiable viewpoint planning for long-horizon task visualization.}
The figure compares visual observations produced by different camera strategies across long-horizon tasks.
While previous viewpoint selection methods may capture the overall scene, they often lack semantic awareness of task-relevant regions, leading them to under-emphasize critical parts, subtle state transitions, or evolving relations that are essential for understanding how the task unfolds.
VISTA instead adapts the viewpoint over time to highlight the most informative object parts and interactions, producing more legible and functionally meaningful task visualizations.}
    \label{fig:teaser}
\end{figure*}

Visualization and illustration have long been effective at conveying how things look. A rich body of work~\cite{zhu2011sketch,tatzgern2014hedgehog,Hsu2013} has developed techniques for depicting shape, appearance, spatial structure, and static relationships. By contrast, helping people understand functional behavior and state changes remains substantially more challenging~\cite{Mitra2013,li2004interactive}. To communicate a process, a mechanism, or a long-horizon task, it is not enough to show the scene clearly at isolated moments. Viewers must be able to follow how relevant parts change over time, how actions transform states, and how local interactions accumulate into meaningful outcomes. In other words, ``how-things-work'' visualization requires not only visibility, but also a coherent visual narrative that unfolds progressively over time.

\highlight{In this paper, we use \emph{long-horizon tasks} to refer to temporally extended object-interaction processes composed of multiple semantically meaningful action steps. Such tasks require tracking how object parts, spatial relations, and task-relevant states evolve over time, rather than interpreting each moment in isolation~\cite{puig2018virtualhome,srivastava2021behavior,mees2022calvin}. We specifically focus on \emph{object-centered long-horizon tasks} in interactive 3D environments, rather than broader scenarios involving full human activity understanding, social interaction, or whole-scene navigation.} Prior work has demonstrated the value of computational methods for illustrating operation rather than static form alone, such as depicting how mechanical assemblies function, how fluid systems behave, or how geological structures evolve. These works highlight an important shift in emphasis: from showing what an object is to explaining what it does. Yet despite this progress, modern 3D visualization and simulation systems still provide limited support for constructing such process-oriented explanations. In many systems, camera control is treated as a fixed rendering choice~\cite{hartwig2022learning,gao2025realappliancelethighfidelityappliance}, a manually authored animation path~\cite{andujar2004wayfinder}, or a viewpoint optimization problem driven primarily by geometry~\cite{neumann2005viewpoint}.

While existing approaches may be adequate for short and atomic events, they become increasingly insufficient for long-horizon processes, defined here as multi-step processes that unfold over extended time spans and require viewers to track state changes, action dependencies, and spatially distributed interactions across successive stages. As tasks unfold, relevant parts may become too small, occluded, or drift outside the field of view. More fundamentally, viewpoints chosen only for instantaneous visibility often fail to preserve the continuity needed for understanding a process step by step. They may reveal what is visible in the current frame, yet fail to show what changed, why it changed, and how the current state relates to previous actions. As a result, viewers may perceive the scene but do not fully understand underlying mechanisms.  \highlight{This limitation is especially pronounced in object-centered manipulation tasks involving articulated or stateful objects, where an agent or system performs a sequence of actions that produces observable state changes over time. We use \emph{embodied tasks} to refer to tasks situated in physical or simulated 3D environments, where an agent interacts with articulated or stateful objects, such as appliances, furniture, vehicles, and other household artifacts. In the agent-facing setting, these tasks become sequential decision-making problems: each action depends on visual observations of the current object state, the task goal, and the history of previous interactions. This scope is consistent with embodied AI and robotics benchmarks that model household activities as sequences of atomic actions and object-state transitions~\cite{puig2018virtualhome,shridhar2020alfred,srivastava2021behavior}, articulated-object environments requiring part-level interaction~\cite{xiang2020sapien}, and long-horizon robot manipulation benchmarks~\cite{james2020rlbench,mees2022calvin,gu2023maniskill2}.}

We argue that camera control for long-horizon task visualization should be treated as a visualization policy. From this perspective, the camera is not merely a rendering parameter or a cinematographic afterthought; it is an active mechanism for organizing a linear visual narrative of the process. A useful camera policy  should therefore help viewers follow the evolution of task-relevant states, maintain sufficient structural context across steps, and preserve the causal relationship between actions and outcomes~\cite{Segel2010NarrativeVisualization,Heer2007AnimatedTransitions,Zacks2007EventSegmentation}. Rather than optimizing only local visibility or aesthetic smoothness, it should support a temporally coherent visual explanation of how the process unfolds.

In this work, we present a differentiable viewpoint planning framework for long-horizon task visualization. Our method directly optimizes camera parameters under a unified projection-based quality objective. To support informative local views without losing global context, we introduce a state-aware formulation that jointly considers the target part, its parent structure, and the semantics of the ongoing action. The resulting planner produces adaptive viewpoints that evolve with the process while remaining temporally coherent. We demonstrate the versatility of our approach in two complementary settings. In a presentation-driven setting, given a long-horizon action sequence, the planner generates a camera trajectory and animation that visually narrate the task for human viewers, helping them understand how the process unfolds over time. In an agent-observation setting, the same planner provides a vision language model(VLM) agent with more informative viewpoints during observation, improving the visual evidence available for decision making in embodied tasks. These two settings highlight that our method supports both human-facing process understanding and agent-facing embodied decision making. To validate the perceptual quality of our generated camera trajectories, we conducted a user study with 21 participants comparing our method against a fixed camera baseline and an entropy-based approach. Our method was rated significantly higher across all evaluation criteria and was ranked first in 96\% of all tasks. 
We show that camera viewpoint is a critical yet underexplored factor in VLM-based embodied tasks, and propose a benchmark (\textbf{VISTA-Bench}) that systematically evaluates the impact of viewpoint across perception (task 1), goal reasoning (task 2), and closed-loop control (task 3).
Across diverse appliances and state-based tasks, our approach improves perception accuracy, goal decision, and, most notably, closed-loop control, achieving up to \textbf{$2.1\times$} higher success rates. 
These results highlight that adaptive viewpoint selection is essential for reliable sequential decision-making in embodied AI.

\highlight{
Our contributions are fourfold:
\begin{itemize}

\item We formulate viewpoint planning for object-centered long-horizon tasks around action-conditioned visual targets. Rather than selecting a generic best view of an entire object, each action instantiates either an object--part pair or a relational tuple, allowing different steps on the same object to emphasize different manipulated
parts, interacting entities, and contextual structures.

\item We translate these task-conditioned requirements into interpretable differentiable image-space objectives. In particular, we combine five metrics that jointly account for visibility, semantic relevance, and temporal coherence, enabling task-relevant visual criteria to be optimized directly with respect to continuous camera parameters.

\item We implement the proposed metrics in VISTA, an interactive prototype for differentiable viewpoint planning. Through its graphical interface and configuration files, users can adjust the metric weights, task-importance distributions, target coverage, camera constraints, and optimization settings, enabling the planner to be configured for different objects, actions, and visualization requirements.

\item We evaluate the VISTA for both human-facing narrative visualization and VLM-based active observation, and introduce VISTA-Bench to isolate the effect of viewpoint quality on perception, goal reasoning, and sequential control.
\end{itemize}
}

\section{Related Work}
\label{sec:relatedwork}

\subsection{Heuristic View Quality Measures}
A substantial body of prior work studies viewpoint selection through hand-crafted heuristics and quality measures for \emph{static} 3D objects. Early methods rely on geometric cues derived from rendered views, such as visible polygons and projected area~\cite{Bonaventura2018SurveyViewpointSelection,Vazquez2001ViewpointEntropy,Vazquez2003AutomaticViewSelection,neumann2005viewpoint}, silhouette shape~\cite{vieira2009learning,page2003shape,polonsky2005s}, depth variation~\cite{stoev2002case,Secord2011PerceptualViewpointPreference}, and saliency-related cues~\cite{Lee2005MeshSaliency}. Beyond geometry, prior work has also discussed viewpoint preference in terms of canonicality, familiarity, functionality, and aesthetics~\cite{dutagaci2010benchmark,palmer1981cannonical,koulieris2014c,becker2007object}, suggesting that a ``good'' view depends not only on visibility but also on how an object is commonly perceived or used. In practice, however, most classical computational methods still optimize quantities that can be directly measured from geometry or rendering, such as visible area, contour complexity, saliency, or information content.

Among these measures, \emph{viewpoint entropy} is one of the most influential information-theoretic formulations. It measures how evenly the projected areas of visible polygons are distributed in the image and prefers views that expose many parts of the object in a balanced way~\cite{Vazquez2001ViewpointEntropy,Vazquez2003AutomaticViewSelection}. Using the standard notation over polygons $z\in Z$, viewpoint entropy is defined as
\begin{equation}
VE(v) = - \sum_{z\in Z} \frac{a_z(v)}{a_t(v)}
\log \frac{a_z(v)}{a_t(v)}
= - \sum_{z\in Z} p(z\mid v)\log p(z\mid v),
\end{equation}
where $a_z(v)$ is the projected area of polygon $z$ from viewpoint $v$, $a_t(v)$ is the total projected area of the model, and $p(z\mid v)$ is the normalized projected-area distribution. High entropy corresponds to views that show many polygons with relatively balanced visibility.

A related but more expressive measure is \emph{viewpoint mutual information} (VMI)~\cite{Viola2006ImportanceDriven}, which evaluates the correlation between a viewpoint and the object's polygon distribution. In its common form, VMI can be written as
\begin{equation}
VMI(v) = \sum_{z\in Z} p(z\mid v)\log \frac{p(z\mid v)}{p(z)},
\end{equation}
where $p(z)$ is the average projected-area distribution of polygon $z$ over viewpoints. Intuitively, VMI compares the current viewpoint-dependent visibility distribution against a reference distribution and favors representative views with lower coupling to only a small subset of polygons. It can therefore also be interpreted as a KL-style divergence~\cite{neumann2005viewpoint} between the visibility distribution under the current view and a target distribution over object parts.

\highlight{Similar heuristic principles have been extended to volume visualization. Bordoloi and Shen~\cite{Bordoloi2005ViewSelection} propose a view-selection method for volume rendering that favors views where important voxels are highly visible, using visibility and importance to identify informative views or a compact set of representative views. Tao \etal~\cite{Tao2009StructureAware} further introduce structure-aware viewpoint selection for volume visualization, defining shape and detail view descriptors to capture global structures and local boundary details from volume data. These methods broaden static view selection from surface visibility to volumetric visibility, importance, and structural characteristics.}

Despite their importance, heuristic viewpoint selection methods remain limited for our setting. Most are designed to select a single best or representative view of a static object through dense sampling and repeated rendering over candidate viewpoints, making best-view search costly. Moreover, many measures emphasize one dominant cue at a time: geometric methods focus on visible area, contour, or depth; information-theoretic measures such as VE and VMI rely on projected visibility distributions~\cite{Viola2006ImportanceDriven}; and higher-level criteria such as familiarity or functionality are difficult to encode in a unified optimization framework. As a result, a view may score highly while failing to emphasize semantically critical regions relevant to the current task. Although VMI is more stable than viewpoint entropy and can incorporate object-level importance, it still operates on static visibility distributions and does not explicitly model action phases, interaction semantics, or temporal continuity. Overall, these methods provide valuable geometric and information-theoretic foundations, but they remain single-view quality criteria rather than policies for organizing long-horizon visual explanations. Our setting instead requires viewpoint planning for \emph{long-horizon task visualization}, where the camera must jointly account for semantic focus, structural context, and temporal coherence over time. \highlight{In addition, while some prior work focuses on volume data, our work primarily targets mesh-based objects and their task-driven state transitions.}

\subsection{Learning-Based Viewpoint Selection}

Learning-based viewpoint selection has recently emerged as an alternative to brute-force evaluation of handcrafted quality measures. Schelling \etal\ propose a dynamic-label-generation framework that predicts high-quality viewpoints directly from 3D models, reducing the reliance of traditional measures on clean surface meshes and costly rendering-based search~\cite{Schelling2021DynamicLabelGeneration}. Their method also addresses the ambiguity of non-unique optimal views, such as those caused by object symmetries, by adapting the supervision target during training. Another line of work learns viewpoint quality from \emph{human preferences} rather than predefined geometric criteria. Hartwig \etal\ show that preferred views depend not only on engineered measures such as entropy or occupancy, but also on category knowledge and familiar depictions, and train a neural view-quality model from large-scale crowdsourced pairwise comparisons~\cite{hartwig2022learning}.

\highlight{Related learning-based prediction has also been studied in interactive volume visualization. Sun \etal~\cite{Sun2025RmdnCache} propose RmdnCache, a dual-space RNN--MDN model that predicts the next POV and its uncertainty range from user exploration trajectories for data prefetching in large-scale volume rendering. However, its goal is to reduce cache misses and input latency, rather than to estimate viewpoint quality or plan informative viewpoints.} Overall, these works improve the efficiency and flexibility of view prediction, but they still focus on static best-view estimation, preference modeling, or system-level prefetching. In contrast, our work formulates viewpoint selection as a differentiable long-horizon planning problem, where the camera must reveal state transitions, preserve context across actions, and support action--outcome attribution over time.

\subsection{Dynamic Camera Planning and Visibility-Aware Control}

Beyond static viewpoint selection, prior work has explored camera control as a task-driven and optimization-based process. \highlight{Ji~\etal~\cite{Ji2006DynamicViewSelection} propose dynamic view selection for time-varying volumes by first evaluating static view quality using opacity, color, and curvature distributions, and then applying dynamic programming to select a smooth viewing path that maximizes perceived information over time.} Similarly, Mitra~\etal~\cite{Mitra2013} generate visual explanations of dynamic processes, such as exploded views and step-by-step assembly visualizations, emphasizing functional behavior and action sequences rather than purely geometric visibility.

The computer graphics and animation community has also developed formal models for camera control. Christie~\etal~\cite{Christie2009} survey camera control techniques that formulate camera placement and motion as constraint- and optimization-driven problems, balancing visibility, framing, smoothness, and narrative constraints. Hsu~\etal~\cite{Hsu2013} similarly optimize camera trajectories for volume data animation under multiple criteria, including feature visibility, importance, and temporal smoothness. These works show that camera planning is inherently multi-objective.

However, existing approaches typically target short sequences, predefined animations, or fixed optimization objectives. They do not explicitly formulate camera control as a continuous, differentiable policy over long-horizon tasks. In contrast, our work casts viewpoint planning as a unified differentiable optimization problem that encodes task-dependent visualization goals and produces temporally coherent viewpoint sequences for evolving processes.

\section{Design Goals and Unified Camera Quality Objective}\label{design goals}

\highlight{Long-horizon task visualization requires the camera to do more than merely keep a target visible, as illustrated in ~\autoref{fig:overview}. As a task unfolds, the viewpoint must adapt to changing explanatory demands: it should make the task-relevant component and its state changes clearly legible, preserve sufficient surrounding structure for those changes to be interpreted, and maintain visual continuity across successive action stages ~\cite{Segel2010NarrativeVisualization,
Heer2007AnimatedTransitions,
he2026bridging,
liu2025aligningcyberspacephysical}. We therefore formulate camera control as a visualization policy over a long-horizon action sequence. For each action, the policy constructs an action-conditioned visualization target and optimizes continuous camera parameters according to explicit viewpoint-quality objectives.}

\subsection{Action-Conditioned Visualization Targets}
\begin{figure*}[t]
    \centering
    \includegraphics[width=1.0\linewidth]{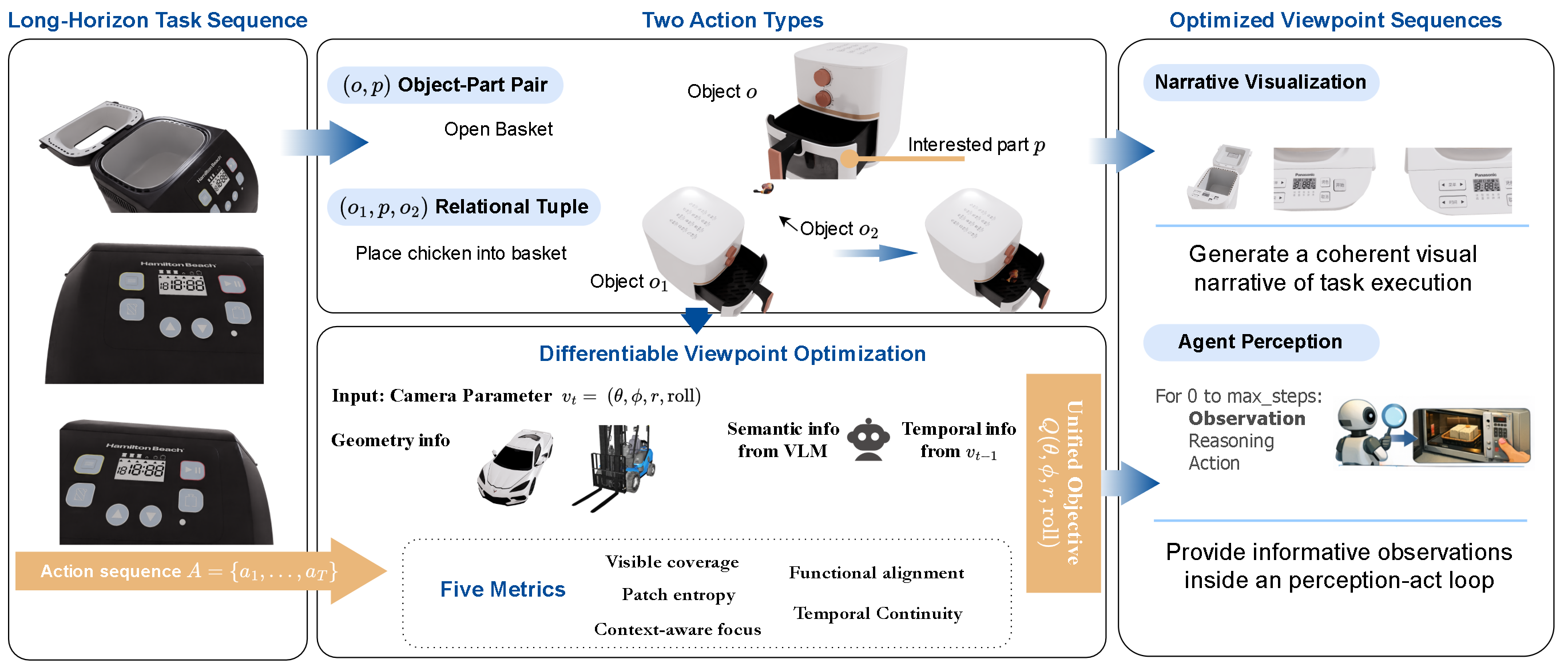}
    \vspace{-4mm}
    \caption{
\textbf{Overview of VISTA.}
Given a long-horizon task sequence, we represent interactions using two complementary formulations: object–part pairs $(o,p)$ for local manipulations and relational tuples $(o_1,p,o_2)$ for interactions involving multiple objects.
At each step, VISTA optimizes camera parameters $v_t$ via a differentiable viewpoint planning module that integrates geometric cues, semantic signals from a VLM, and temporal context.
A unified objective, composed of five metrics (visible coverage, patch entropy, context-aware focus, functional alignment, and temporal continuity), guides the selection of task-relevant viewpoints over time.
The resulting viewpoint sequence supports two applications: (i) narrative visualization that produces coherent, interpretable visual summaries of task execution, and (ii) agent perception that provides informative observations within a closed-loop perception–action pipeline.
}
    \label{fig:overview}
\end{figure*}

To capture these requirements, we formulate adaptive viewpoint planning over an input action sequence $A=\{a_1,\dots,a_T\}$ describing a long-horizon task. Since different actions impose different visualization demands, we consider two common target configurations in long-horizon tasks, illustrated in the middle panel of~\autoref{fig:overview}. The first type is an \emph{object--part} target $(o,p)$, where $p$ is a salient part of object $o$ and the main goal is to clearly inspect a local manipulation or state change on that part. The second type is a \emph{relational} target $(o_1,p,o_2)$, where part $p$ of object $o_1$ interacts with, affects, or should be interpreted relative to another object $o_2$. This type is used when understanding the action requires not only the manipulated part itself, but also its spatial or functional relation to another object. These two configurations cover the primary cases considered in this work, but are not intended as an exhaustive taxonomy of all possible task targets.

\highlight{For each action $a_t$, our goal is to determine a camera configuration $\mathbf{v}_t$ that best supports the corresponding visualization target. The selected view should make the target legible, preserve sufficient object-level context, and, when applicable, reveal the relation between the involved entities. Optimizing one viewpoint for each action produces a temporally ordered sequence that supports both local part inspection and interaction-centric visualization.}

These requirements are difficult to satisfy with existing viewpoint selection paradigms. Hand-crafted heuristic methods~\cite{Vazquez2001ViewpointEntropy,Viola2006ImportanceDriven} typically optimize fixed geometric cues and \highlight{commonly rely on a discrete candidate-search procedure: they sample a large number of viewpoints over the camera space, render and evaluate each candidate independently, and select the one with the highest score.} More importantly, both heuristic~\cite{Vazquez2001ViewpointEntropy} and learning-based methods~\cite{hartwig2022learning} have largely focused on finding the best view of a whole object, rather than a view tailored to a task-relevant part or inter-object relation. This makes them poorly suited to long-horizon tasks, where visualization needs change across action phases: one step may require a close-up of a manipulated part, while the next may require a wider view to reveal its effect on surrounding structure or another object. Learning-based approaches can adapt more flexibly, but they depend on collected training data, and obtaining labels with sufficient spatial coverage over the viewpoint space is expensive and difficult. Their behavior is therefore tied to the training distribution, with limited control over which visualization criteria are actually being optimized. In our setting, we instead seek a planner that can adapt to each task state, remain directly steerable by explicit visualization goals, and optimize views for task-relevant parts and relations rather than whole-object preference.

\highlight{These limitations motivate a differentiable formulation of viewpoint planning. Instead of discretizing the camera space and exhaustively evaluating a predefined set of candidate views, we express viewpoint quality using image-space objectives and optimize continuous camera parameters directly through gradients. This allows the optimizer to progressively adjust the camera toward a better view without requiring dense enumeration of the full viewpoint space. Equally importantly, each visualization requirement remains an explicit and interpretable objective, allowing us to specify which properties a good view should satisfy. The resulting formulation therefore combines continuous camera optimization with direct control over the task-dependent criteria being optimized, rather than relying on fixed camera rules or expecting explanatory behavior to emerge implicitly from training data.}

\highlight{
We formulate adaptive camera control as a visualization policy that selects a sequence of camera configurations
\begin{equation}
V=\{\mathbf{v}_t\}_{t=1}^{T}, \qquad
\mathbf{v}_t=(\theta_t,\phi_t,r_t,\mathrm{roll}_t),
\label{eq:cam}
\end{equation}
where $\theta_t$ and $\phi_t$ denote the azimuth and elevation around the action-dependent camera anchor $\mathbf{c}_{a_t}$, $r_t$ denotes the viewing distance, and $\mathrm{roll}_t$ denotes the in-plane camera rotation around the viewing axis. The objective is therefore not to find one generic best view of an entire object, but to optimize a sequence of viewpoints whose targets and required context evolve with the task.


}
\subsection{Design Goals and Viewpoint-Quality Metrics}

\highlight{We optimize each camera configuration according to five complementary design goals informed by prior work on visibility, structural informativeness, functional perception, importance-driven attention, and animated transitions~\cite{barral2000visual,Vazquez2001ViewpointEntropy,Viola2006ImportanceDriven,Secord2011PerceptualViewpointPreference,harman1999active,Heer2007AnimatedTransitions,Zacks2007EventSegmentation}. Rather than defining these goals separately from their mathematical realizations, we introduce each goal together with the metric used to operationalize it.}

\begin{figure}[tb]
    \centering
    
    \begin{subfigure}[b]{0.48\linewidth}
        \centering
        \includegraphics[width=\linewidth]{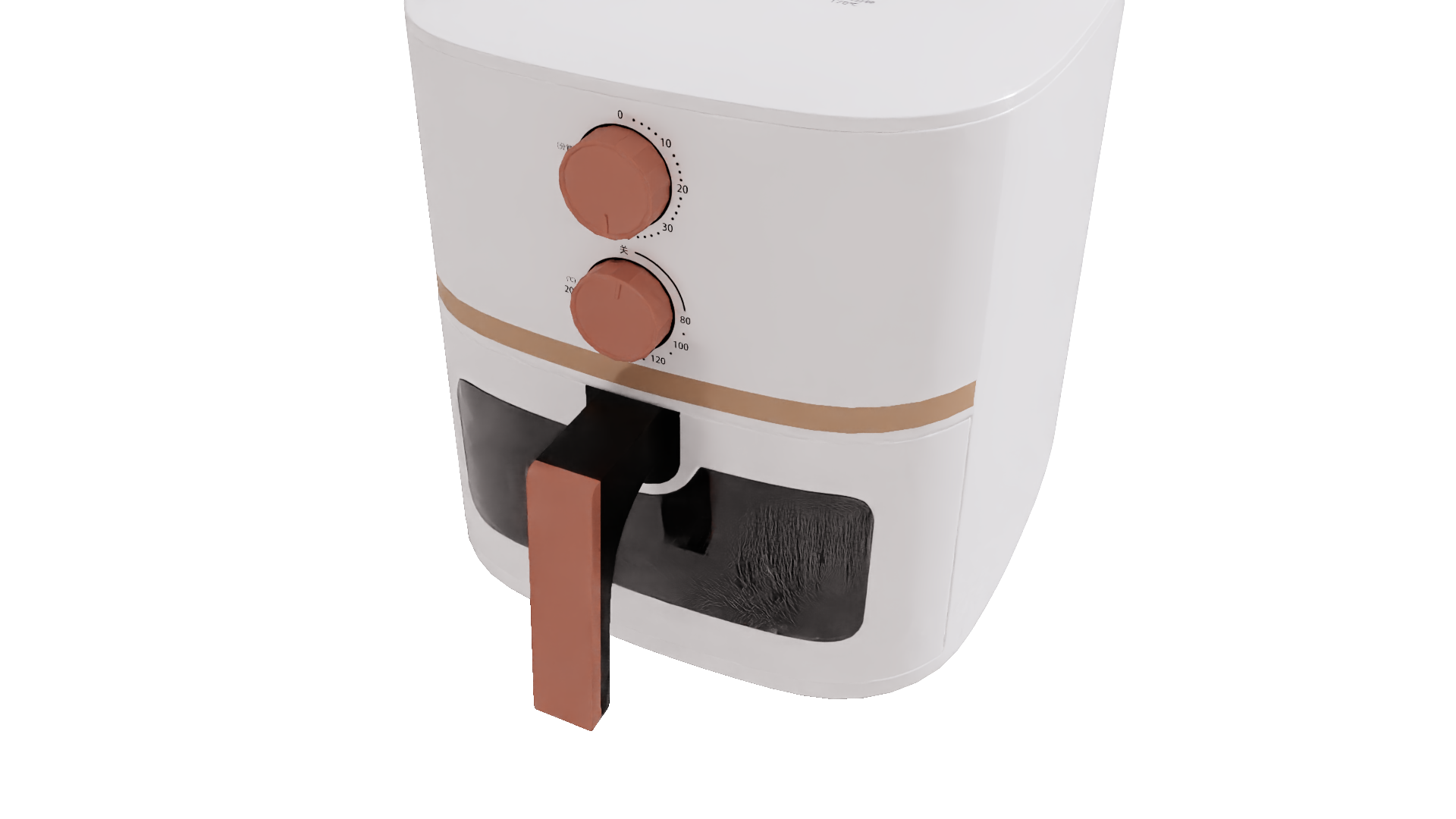}
        \caption{Rendered Image}
        \label{fig:rendered image}
    \end{subfigure}
    \hfill
    \begin{subfigure}[b]{0.48\linewidth}
        \centering
        \includegraphics[width=\linewidth]{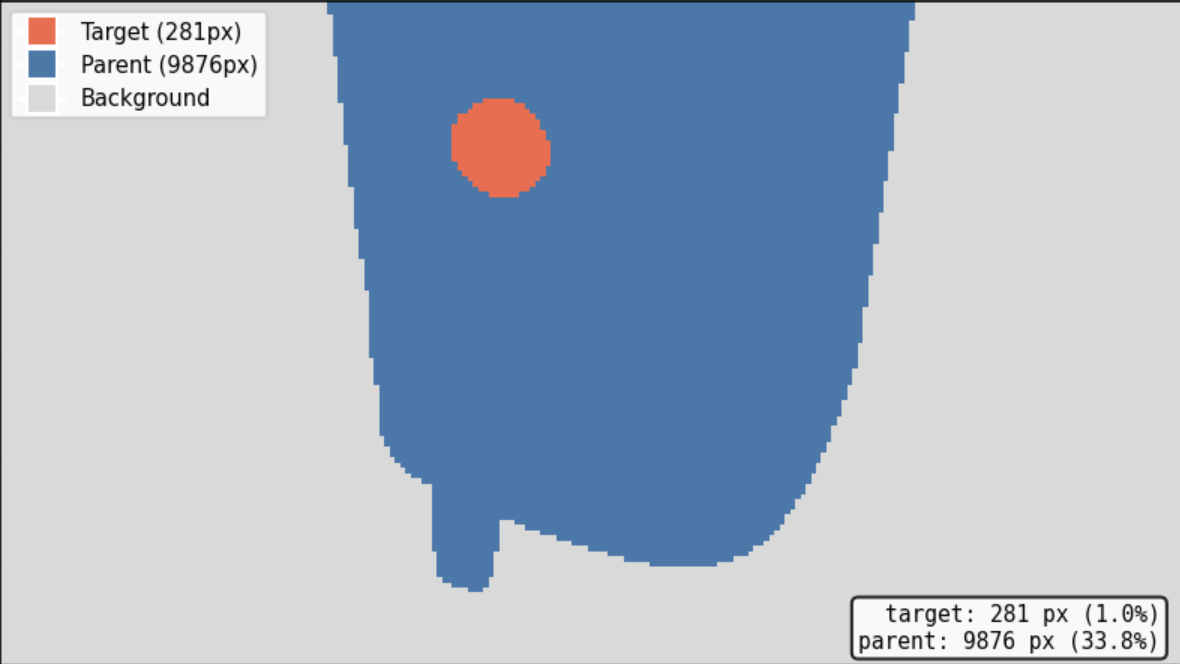}
        \caption{Visible Coverage}
        \label{fig:visible}
    \end{subfigure}

    \vspace{0.5em}

    \begin{subfigure}[b]{0.48\linewidth}
        \centering
        \includegraphics[width=\linewidth]{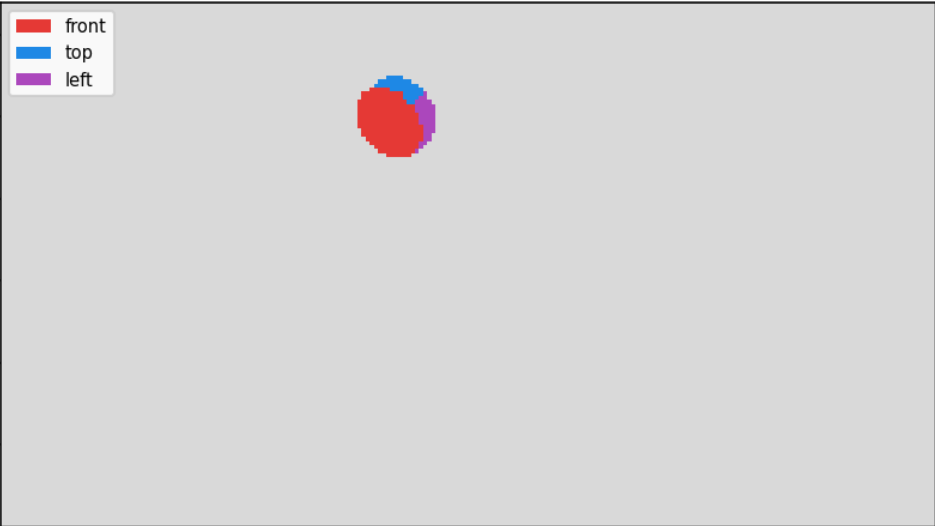}
        \caption{Patch Entropy}
        \label{fig:part}
    \end{subfigure}
    \hfill
    \begin{subfigure}[b]{0.48\linewidth}
        \centering
        \includegraphics[width=\linewidth]{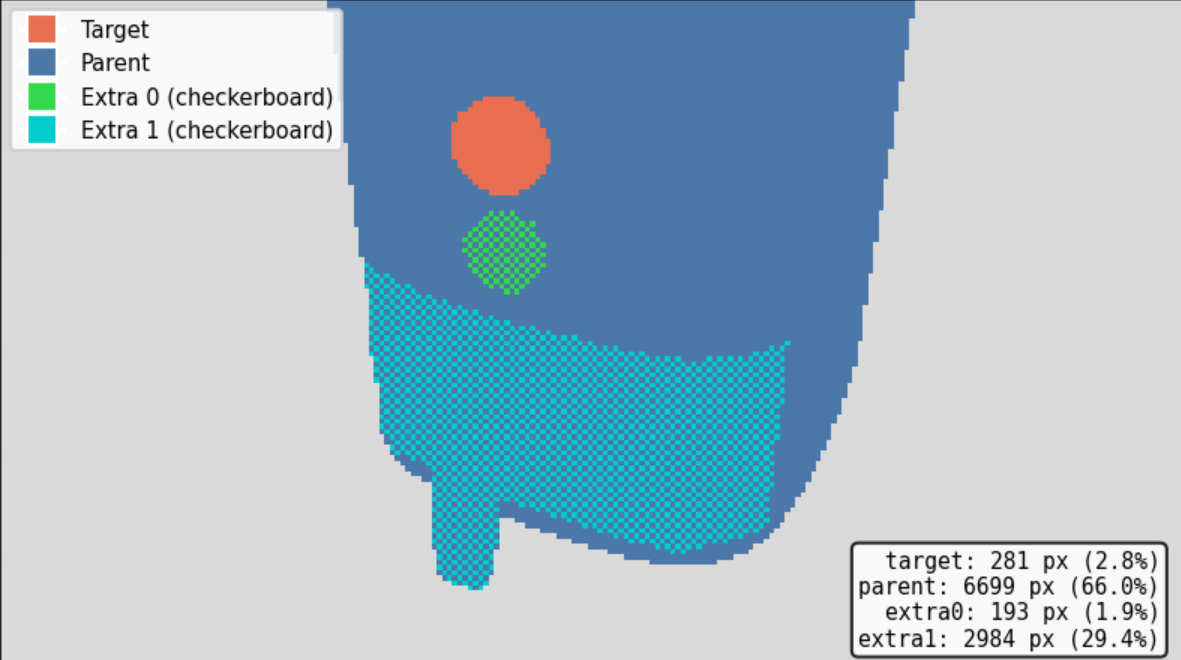}
                \caption{Context-aware focus}
        \label{fig:important}
    \end{subfigure}

    \caption{\highlight{
\textbf{
Visualization of differentiable viewpoint quality metrics for the action \emph{rotate(top knob)}.}
(a) Rendered image of the whole object. For this action, the top knob is the task-relevant target, while the lower knob is not involved.
(b) \emph{Visible coverage}: highlights visible pixels belonging to the target top knob (orange) and its parent object (blue), measuring the proportion of the task-relevant target observed in the image. The lower knob is not highlighted because it is not part of the current action.
(c) \emph{Patch entropy}: visualizes the partition of the target surface into six semantic directional groups defined by the selected candidate-axis frame, encouraging the view to reveal multiple sides of the manipulated part rather than a single flattened surface.
(d) \emph{Context-aware focus}: incorporates pixels from the target, its parent, and surrounding regions, capturing the alignment between the observed visual context and the focus required to perform the action.
Together, these metrics quantify complementary aspects of viewpoint quality, including target visibility, intra-part information diversity, and task-relevant context alignment.
}}
\label{fig:visualization}
\end{figure}

\textbf{G1: Target Legibility - Visible Coverage ($f_{\mathrm{vis}}$):}
Prior work on viewpoint quality has shown that visibility and projected size are central factors in whether a view is informative and perceptually useful ~\cite{barral2000visual,Secord2011PerceptualViewpointPreference}. In our setting, however, the relevant unit is often not the whole object but the task-relevant part on which an action or state change occurs. We therefore measure whether this target is visible, unoccluded, and sufficiently large in the rendered image.
\begin{equation}
f_{\mathrm{vis}}(\mathbf v)=\frac{1}{HW}\sum_{(i,j)\in\Omega}m_{\mathrm{vis}}(i,j;\mathbf v),
\end{equation}
where $\Omega$ is the $H\times W$ image grid, and $(i,j)\in\Omega$ denotes a pixel location. The binary mask $m_{\mathrm{vis}}(i,j;\mathbf v)\in\{0,1\}$ equals $1$ if the target is visible and unoccluded at pixel $(i,j)$ under viewpoint $\mathbf v$, and $0$ otherwise. Thus, $f_{\mathrm{vis}}(\mathbf v)$ measures measures the fraction of the rendered image occupied by visible target content. A higher value indicates that the target is more prominently presented, whereas the score decreases when the target becomes smaller, moves outside the field of view, or is occluded by surrounding geometry.

\highlight{
\textbf{Semantic Directional Frame:}
To represent the target from functionally meaningful directions, we construct a semantic directional frame that captures its front-back, top-bottom, and side-to-side orientations. This frame provides a shared directional reference for evaluating both the structural diversity revealed by a viewpoint and its alignment with the target's functional orientation. We first construct three orthogonal candidate axes using one of two supported policies. The \emph{AABB} policy uses the three world-aligned axes of the target's axis-aligned bounding box, while the \emph{PCA OBB} policy~\cite{dimitrov2006bounding} estimates three object-oriented axes from the target mesh using area-weighted PCA. Both policies produce six opposing candidate directions $\{\pm\mathbf{a}_1,\pm\mathbf{a}_2,\pm\mathbf{a}_3\}$.

We render the target from these six directions and provide the direction-labeled images to a VLM, which identifies the functional front and top directions. For an individual part, each candidate direction is represented by two images: an isolated rendering showing only the part and an in-context rendering in which the part is highlighted within the complete parent object. The isolated views expose local geometry, while the contextual views help the VLM determine how the part is positioned and typically approached during interaction. For each candidate direction, the camera is placed on the corresponding ray from the target's bounding-box center at twice the target bounding-box diagonal for isolated views and 1.2 times the parent bounding-box diagonal for contextual views.

Let $\hat{\mathbf{d}}_{\mathrm{front}}$ and
$\hat{\mathbf{d}}_{\mathrm{up}}$ denote the resulting target-level front and up directions. After orthogonalizing the predicted up direction with respect to the front direction, we construct a side direction and obtain the semantic frame
$\{\pm\mathbf{d}_f,\pm\mathbf{d}_u,\pm\mathbf{d}_s\}$
where $\mathbf{d}_f$, $\mathbf{d}_u$, and $\mathbf{d}_s$ denote the semantic front, up, and side axes, respectively.

This semantic frame serves two complementary purposes in the following metric definitions: it provides the directional partition used to measure the structural diversity revealed by a viewpoint, and it provides the functional front and upright references used to evaluate camera alignment. Detailed prompts of VLM are provided in the Supplementary Material-Section 1.
}

\highlight{
\textbf{G2: Structural Informativeness --- Patch Entropy
($f_{\mathrm{ent}}$):} Prior work on viewpoint entropy and related informativeness measures suggests that informative viewpoints should reveal meaningful structural evidence of the underlying 3D object rather than collapse it into a degenerate or uninformative projection~\cite{Vazquez2001ViewpointEntropy}. For long-horizon task visualization, this means that the selected view should preserve sufficient structural evidence of the task-relevant target for its three-dimensional form and state to remain interpretable.

Classical viewpoint entropy measures structural informativeness by computing entropy over visible polygons~\cite{Vazquez2001ViewpointEntropy}. However, polygon-level formulations are sensitive to mesh triangulation~\cite{neumann2005viewpoint} and may primarily reflect geometric discretization rather than task-relevant surface diversity. We therefore compute entropy over a fixed vocabulary of six semantic directional patches, front, back, top, bottom, left, and right, defined by the semantic frame above.

Each mesh triangle $f$ is assigned to one directional group according to
\begin{equation}
s(f,m)=
\alpha\langle\mathbf n_f,\mathbf d_m\rangle+
(1-\alpha)\langle\hat{\mathbf c}_f,\mathbf d_m\rangle,
\label{eq:main-patch-assignment}
\end{equation}
where $\mathbf n_f$ is the outward triangle normal, $\hat{\mathbf c}_f$ is the normalized triangle-centroid location, $\mathbf d_m$ is the $m$-th canonical direction, and $\alpha=0.5$. The orientation term groups similarly facing surfaces, while the location term disambiguates curved or oblique surfaces. All triangles assigned to the same direction form one patch. 

For viewpoint $\mathbf v$, let
\begin{equation}
A_n(\mathbf v)=
\sum_{(i,j)\in\Omega}m_n(i,j;\mathbf v)
\end{equation}
denote the visible pixel area belonging to patch $P_n$, where
$m_n(i,j;\mathbf v)$ is the visibility indicator of patch $P_n$ at
pixel $(i,j)$. We define
\begin{equation}
\pi_n(\mathbf v)=
\frac{A_n(\mathbf v)}
{\sum_{n'=1}^{N}A_{n'}(\mathbf v)}
\end{equation}
and compute normalized patch entropy as
\begin{equation}
f_\mathrm{ent}(\mathbf v)=
-\frac{1}{\log N}
\sum_{n=1}^{N}
\pi_n(\mathbf v)\log\pi_n(\mathbf v).
\end{equation}

Full implementation details and comparisons with alternative patch assignment policies are provided in the Supplementary Material Section 2.}

\highlight{
\textbf{G3: Functional Alignment
($f_{\mathrm{func}}$):}
Studies on active and task-dependent perception suggest that object understanding benefits from viewpoints selected relative to the observer's goal and the functional properties of the object, rather than from generic canonical views alone~\cite{harman1999active}. In our setting, some directions are more informative because they better reveal how a part is normally observed or operated, such as the front of a button, the readable face of a dial, or the operational side of a handle.

Let $\hat{\mathbf d}_\mathrm{front}$ and
$\hat{\mathbf d}_\mathrm{up}$ denote the resulting target-level front
and up directions. Let $\hat{\mathbf c}(\mathbf v)$ denote the unit
direction from the target toward the camera, and let
$\hat{\mathbf u}_\mathrm{cam}(\mathbf v)$ denote the camera up vector.
We define
\begin{equation}
f_\mathrm{func}(\mathbf v)=
\max\!\bigl(
\langle\hat{\mathbf c}(\mathbf v),
\hat{\mathbf d}_\mathrm{front}\rangle,0
\bigr)
\cdot
\max\!\bigl(
\langle\hat{\mathbf u}_\mathrm{cam}(\mathbf v),
\hat{\mathbf d}_\mathrm{up}\rangle,0
\bigr).
\label{eq:front}
\end{equation}

The first factor favors cameras located on the functional front side of the target, while the second maintains an upright and semantically interpretable image orientation. Although Functional Alignment and Patch Entropy use the same semantic frame, they capture different properties. Functional Alignment determines whether the camera follows a functionally meaningful direction, whereas Patch Entropy evaluates whether the viewpoint reveals multiple directional surface regions of the target.

}
\highlight{
\textbf{G4: Context-Aware Focus ($f_{\mathrm{foc}}$):}
Prior work on importance-driven focus of attention suggests that a viewpoint should allocate image-space emphasis according to the relative importance of scene elements rather than treat all visible content equally~\cite{Viola2006ImportanceDriven}. In long-horizon tasks, however, importance depends not only on individual objects, but also on the contextual information needed to interpret the manipulated target and its relation to the surrounding scene.

We therefore measure whether the current viewpoint allocates image space among the target and its surrounding entities according to their task-dependent importance. The objective is not to maximize target area alone: an excessively close view may remove the context needed to interpret the target's role, whereas a view containing too much surrounding structure may make the target too small to inspect.

For each entity $e\in E$, let
\begin{equation}
A_e(\mathbf v)=\sum_{(i,j)\in\Omega} m_e(i,j;\mathbf v)
\end{equation}
denote the number of rendered pixels occupied by $e$. We then define the visual distribution induced by the current view as
\begin{equation}
p(e\mid \mathbf v)=\frac{A_e(\mathbf v)}{\sum_{e'\in E}A_{e'}(\mathbf v)}.
\end{equation}
Let $q(e)$ denote a task-importance distribution specifying how visual attention should ideally be allocated across entities at the current action step. The context-aware focus score is defined as
\begin{equation}
f_\mathrm{foc}(\mathbf v)=
\exp\!\bigl(-D_\mathrm{JS}(p(\cdot\mid\mathbf v)\,\|\,q(\cdot))\bigr),
\end{equation}
with
\begin{equation}
D_\mathrm{JS}(p\|q)=
\frac{1}{2}D_\mathrm{KL}(p\|m)+\frac{1}{2}D_\mathrm{KL}(q\|m),
\qquad
m=\frac{1}{2}(p+q),
\end{equation}
with $D_\mathrm{KL}$ being the Kullback-Leibler divergence.

This score is high when the current view allocates screen space across the target and its context in a way that matches their task-dependent importance. We use Jensen--Shannon divergence because it provides a symmetric and bounded measure of distribution mismatch, which is more suitable than a single KL divergence for comparing task-dependent attention allocations.

}
\highlight{
\textbf{G5: Temporal Continuity ($f_{\mathrm{trans}}$):}
Prior work on animated transitions and event understanding has shown that coherent visual changes help viewers preserve spatial context, maintain orientation, and follow evolving processes over time~\cite{Heer2007AnimatedTransitions,Zacks2007EventSegmentation}. For long-horizon task visualization, consecutive camera states should therefore be connected without unnecessarily large changes in viewpoint.
Let $\mathbf{v}_{t-1}$ and $\mathbf{v}_t$ denote the camera configurations associated with two consecutive action steps, and let $\mathbf{p_{t-1}}$ and $\mathbf{p_{t}}$ denote their corresponding camera positions. Let $\ell(\mathbf{p_{t-1}},\mathbf{p_t})$ denote the length of the transition path connecting these positions. We define
\begin{equation}
f_{\mathrm{trans}}
(\mathbf{v}_{t-1},\mathbf{v}_t)
=
\exp\!\left(
-\frac{
\ell(\mathbf{p}_{t-1},\mathbf{p}_t)
}{
d_{\mathrm{ref}}
}
\right),
\label{eq:metric-transition}
\end{equation} where $d_{\mathrm{ref}}>0$ is a reference distance used to normalize the path length. A higher score favors nearby camera configurations connected by shorter viewpoint transitions.

As a proof of concept, we use path length as a lightweight geometric proxy for temporal continuity. This formulation encourages spatially compact transitions but does not directly model higher-order camera-motion or perceptual properties, such as velocity, acceleration, screen-space target displacement, optical flow, or visual correspondence. The visualizations of each metric are listed in~\autoref{fig:visualization}.
}

\section{Image-Space Optimization via Differentiable Rasterization}
\label{sec:dr}

\highlight{
Previous section defines five complementary viewpoint-quality metrics. Among them, Visible Coverage, Patch Entropy, and Context-Aware Focus rely on binary visibility masks, discrete pixel ownership, and projected image-space areas. These rendering quantities are not directly differentiable with respect to the camera parameters: their values may remain unchanged under small camera perturbations and change abruptly when visibility or occlusion boundaries are crossed. Functional Alignment is already expressed as an analytic function of the camera orientation, while Temporal Continuity depends on the transition path between consecutive camera states.

In this section, we construct a shared differentiable camera-to-image mapping for these metrics. Rather than redefining their semantic objectives, we replace hard rendering quantities with soft image-space attributes produced through differentiable projection, rasterization, interpolation, and antialiasing. This allows gradients from the viewpoint-quality objective to propagate directly to the continuous camera parameters.}

\subsection{Differentiable Camera-to-Image Mapping}

We instantiate our viewpoint optimization on top of a standard differentiable rasterization pipeline~\cite{laine2020modular}. Our role here is not to introduce a new renderer, but to define a camera parameterization and a set of task-specific attribute encodings that expose differentiable image-space signals for evaluating candidate views.

The camera position $p(v_t)$ is parameterized using \autoref{eq:campos}

\begin{equation}
\mathbf{p}(\mathbf{v}_t) =
\mathbf{c_{a_t}} +
r_t\begin{bmatrix}
\cos\phi_t\cos\theta_t\\
\cos\phi_t\sin\theta_t\\
\sin\phi_t
\end{bmatrix}.
\label{eq:campos}
\end{equation}
Together with a viewing direction towards target center $\mathbf{c_{a_t}}$ and roll-dependent in-plane rotation, this defines a differentiable view matrix $\mathbf{V}(\mathbf{v_t})$. Let $\mathbf{P}$ denote the perspective projection matrix. For a mesh vertex $\mathbf{x}_i$, the clip-space coordinate is
$\tilde{\mathbf{x}}_i(\mathbf{v}_t) =
\mathbf{P}\,\mathbf{V}(\mathbf{v_t})\,\mathbf{x}_i.$
Since both the camera transform and projection are smooth functions of $\mathbf{v}$, the projected geometry is differentiable with respect to the camera parameters.

\subsection{Soft Rasterized Attributes}
\highlight{
Let $\Omega$ denote the image domain and let $q=(i,j)\in\Omega$ denote a pixel. Each mesh vertex $\ell$ is assigned an attribute vector $\mathbf{a}_{\ell}\in\mathbb{R}^{K}$. For a pixel covered by the front-most triangle $\tau_q=(\ell_0,\ell_1,\ell_2)$, let $\beta_k(q;\mathbf{v}_t)$ denote its perspective-correct barycentric coordinates. The point-sampled per-pixel attribute is
\begin{equation}
\mathbf{a}^{\mathrm{pt}}(q;\mathbf{v}_t) =
\begin{cases}
\displaystyle
\sum_{k=0}^{2}
\beta_k(q;\mathbf{v}_t)\,\mathbf{a}_{\ell_k},
& q \text{ is covered by } \tau_q,\\[5pt]
\mathbf{0},
& \text{otherwise}.
\end{cases}
\label{eq:pixel_attribute}
\end{equation}

We define the soft per-pixel attribute as
\begin{equation}
\tilde{\mathbf{a}}(q;\mathbf{v}_t)
=
\mathcal{A}\!\left(
\mathbf{a}^{\mathrm{pt}}(\cdot;\mathbf{v}_t),
\{\tilde{\mathbf{x}}_{\ell}(\mathbf{v}_t)\},
F
\right)(q),
\label{eq:soft_pixel_attribute}
\end{equation}
where $F$ is the triangle set and $\mathcal{A}$ denotes geometry-aware differentiable antialiasing. The operator approximates pixel coverage at silhouette and visibility boundaries, producing fractional attribute values at boundary pixels while preserving the interpolated attributes in triangle interiors.

For binary target membership, target vertices are assigned $1$ and all other vertices are assigned $0$, so that $\tilde{a}(q;\mathbf{v}_t)\in[0,1]$ represents the fractional target coverage of pixel $q$. For categorical attributes such as entity identity or patch identity, $\mathbf{a}_{\ell}$ is a one-hot vector rather than an integer label. Consequently, $\tilde{\mathbf{a}}(q;\mathbf{v}_t)$ represents the soft per-pixel membership across the corresponding categories.

This construction defines the differentiable mapping
\begin{equation}
\mathbf{v}_t
\;\longrightarrow\;
\{\tilde{\mathbf{x}}_{\ell}(\mathbf{v}_t)\}
\;\longrightarrow\;
\{\tilde{\mathbf{a}}(q;\mathbf{v}_t)\}_{q\in\Omega},
\end{equation}
allowing gradients from image-space objectives to flow back to the camera parameters.
}
 
\highlight{
Visibility is determined by the front-most surface returned by the rasterizer. Because target, entity are interpolated only from the front-most triangle at each pixel, the resulting channels already account for occlusion by the parent object and other scene geometry. Differentiable antialiasing provides fractional coverage near projected boundaries, allowing small camera changes around these boundaries to produce continuous changes in the corresponding image-space attributes.
}

\subsection{Differentiable Realization of the Metrics}

With the soft rasterized attributes defined above, we obtain differentiable counterparts of the three rendering-based metrics introduced in Section~\ref{design goals}. Their semantic definitions remain unchanged; binary masks and hard projected areas are replaced by soft image-space quantities that vary continuously with the camera configuration $\mathbf{v}_t$. Functional Alignment is already an analytic function of the camera orientation, while Temporal Continuity is computed from a differentiable approximation of the transition-path length.

\textbf{Visible coverage}:
We define the visible coverage of the target by integrating the soft visibility mask over the image domain:
\begin{equation}
\tilde{f}_{\mathrm{vis}}(\mathbf{v}_t)
=
\frac{1}{|\Omega|}
\sum_{q\in\Omega}
\tilde{m}_{\mathrm{vis}}(q;\mathbf{v}_t).
\label{eq:dr_vis}
\end{equation}
This term increases when a larger visible portion of the target occupies the image, and decreases under occlusion or shrinkage in projection. In practice, however, maximizing visible coverage without bound is not always desirable, since an excessively large target region may reduce surrounding context. We therefore apply a saturation function
\begin{equation}
\hat{f}_{\mathrm{vis}}(\mathbf{v})
=
\min\!\left(\frac{\tilde{f}_{\mathrm{vis}}(\mathbf{v})}{f_{\mathrm{vis}}^{\mathrm{target}}},\,1\right)
\exp\!\left(-k\,\max\!\bigl(\tilde{f}_{\mathrm{vis}}(\mathbf{v})-f_{\mathrm{vis}}^{\mathrm{target}},\,0\bigr)\right),
\end{equation}
where $f_{\mathrm{vis}}^{\mathrm{target}}$ is a manually specified target coverage. This design encourages the optimizer to maintain sufficient target visibility while avoiding overly zoomed-in views.

\textbf{Patch entropy}:
To measure structural informativeness, we partition the target surface into $N$ semantic or geometric patches. Let $\tilde{w}_n(q;\mathbf{v})\in[0,1]$ denote the soft contribution of patch $n$ at pixel $(i,j)$, obtained by rasterizing patch-identity attributes into image space. The soft projected area of patch $n$ is
\begin{equation}
\tilde{A}_n(\mathbf{v}_t)
=
\sum_{q\in\Omega}\tilde{w}_n(q;\mathbf{v}_t),
\label{eq:dr_area}
\end{equation}
which induces the normalized image-space patch distribution
\begin{equation}
\tilde{\pi}_n(\mathbf{v}_t)
=
\frac{\tilde{A}_n(\mathbf{v}_t)}
{\sum_{n'=1}^{N}\tilde{A}_{n'}(\mathbf{v}_t)}.
\label{eq:dr_pi}
\end{equation}
The differentiable entropy objective is then
\begin{equation}
\tilde{f}_{\mathrm{ent}}(\mathbf{v})
=
-\frac{1}{\log N}
\sum_{n=1}^{N}
\tilde{\pi}_n(\mathbf{v})
\log\!\bigl(\tilde{\pi}_n(\mathbf{v})\bigr),
\label{eq:dr_ent}
\end{equation}
This term favors views that reveal multiple target patches with balanced visibility, thereby increasing structural informativeness.

\textbf{Context-aware focus}:
Let $E$ denote the set of relevant entities, including the target, its parent, and other contextual elements. For each entity $e\in E$, let $\tilde{m}_e(u;\mathbf{v})\in[0,1]$ denote its soft image-space occupancy. The soft projected area of entity $e$ is
\begin{equation}
\tilde{A}_e(\mathbf{v}_t) = \sum_{q\in\Omega}\tilde{m}_e(q;\mathbf{v}_t).
\label{eq:dr_entity_area}
\end{equation}
These areas define a differentiable visual distribution
\begin{equation}
\tilde{p}(e\mid \mathbf{v}_t)
=
\frac{\tilde{A}_e(\mathbf{v}_t)}
{\sum_{e'\in E}\tilde{A}_{e'}(\mathbf{v}_t)}.
\label{eq:dr_focus_dist}
\end{equation}
Given a target importance distribution $q(e)$ determined by task semantics, we compute the focus term by comparing $\tilde{p}(e\mid\mathbf{v})$ with $q(e)$. Using the Jensen--Shannon form introduced earlier, we define
\begin{equation}
\tilde{f}_{\mathrm{foc}}(\mathbf{v})
=
\exp\!\bigl(-D_{\mathrm{JS}}(\tilde{p}(\cdot\mid\mathbf{v}), q(\cdot))\bigr).
\label{eq:dr_foc}
\end{equation}
This term is maximized when the image-space pixel allocation matches the importance distribution of current task state.

\textbf{Temporal continuity}:
To encourage smooth and safe viewpoint changes over time, we define a differentiable transition term between the previous camera position $\mathbf{p}_{t-1}$ and the current one $\mathbf{p}_t$. We represent the camera path as a cubic B\'ezier curve
\begin{equation}
\mathbf{B}(\tau)
=
(1-\tau)^3 P_0
+3(1-\tau)^2\tau P_1
+3(1-\tau)\tau^2 P_2
+\tau^3 P_3,
\label{eq:bezier}
\end{equation}
where $P_0=\mathbf{p}_{t-1}$ and $P_3=\mathbf{p}_t$, and $P_1,P_2$ are fixed intermediate control points determined by a collision-aware path construction.

We measure path efficiency by the arc length
\begin{equation}
\ell
=
\sum_{i=1}^{N}
\bigl\|
\mathbf{B}(\tau_i)-\mathbf{B}(\tau_{i-1})
\bigr\|_2,
\quad \tau_i=\frac{i}{N},
\label{eq:arclength}
\end{equation}

\subsection{Unified Objective and Optimization}

We combine the above terms into a unified viewpoint quality function:
\begin{equation}
Q(\mathbf{v})
=
\tilde{f}_{\mathrm{vis}}(\mathbf{v})^{w_1}
\cdot
\tilde{f}_{\mathrm{ent}}(\mathbf{v})^{w_2}
\cdot
f_{\mathrm{func}}(\mathbf{v})^{w_3}
\cdot
\tilde{f}_{\mathrm{foc}}(\mathbf{v})^{w_4}
\cdot
f_{\mathrm{trans}}(\mathbf{v};\mathbf{v}_{t-1})^{w_5},
\label{eq:quality}
\end{equation}
where $w_i\geq0$ controls the contribution of each design goal. The multiplicative form ensures that severe failure on any one objective, such as complete occlusion of the target, cannot be fully compensated by gains on unrelated terms.

\highlight{
For optimization, we minimize the equivalent negative log-form objective. To account for action dynamics, we sample $K$ intermediate states along the execution of action $a_t$ and evaluate the four per-view metrics under the same camera configuration $\mathbf{v}_t$ at each sampled state. Their losses are averaged across the sampled states, while the temporal-continuity term is evaluated once between $\mathbf{v}_{t-1}$ and $\mathbf{v}_t$. This motion-aware formulation encourages the selected viewpoint to remain informative throughout the action rather than only at a single static instant.
\begin{equation}
\mathcal{L}_t(\mathbf{v}_t)
=
-\frac{1}{K}
\sum_{k=1}^{K}
\sum_{i=1}^{5}
w_i
\log\!\left(
f_i^{(k)}(\mathbf{v}_t)
\right),
\label{eq:loss}
\end{equation}
Here, $f_i^{(k)}$ denotes the $i$-th viewpoint-quality metric evaluated at the $k$-th sampled action state. Since temporal continuity depends only on consecutive camera configurations, its value is shared across all $K$ sampled states.
}

\section{Implementation Details}
\highlight{
We implement VISTA in Isaac Sim 4.5~\cite{nvidia_isaac_sim} as an interactive extension built on the Omniverse Kit framework. The extension integrates the viewpoint planner into the simulator's user interface, rendering pipeline, and articulated-object control system. We evaluate VISTA using mesh-based assets from three sources. RealAppliance~\cite{gao2025realappliancelethighfidelityappliance} contains high-fidelity appliance models with articulated physical and electronic mechanisms and action logic consistent with their manuals. The Chevrolet C8 model~\cite{techfusion_chevrolet_c8_2024} provides a high-fidelity vehicle model with articulated parts.The built-in Isaac Sim asset library provides the articulated Forklift~\cite{nvidia_forklift_c} model, which contains movable components and part-level geometry. Each dataset provides a list of actions supported for each part. To reduce rasterization cost for geometrically complex components, we construct level-of-detail meshes for selected target parts. 

\subsection{Semantic Direction Extraction}

For semantic-face identification, we use a fixed version of \textit{GPT-5.4-2026-03-05}~\cite{openai_gpt54_2026}. The model identifies the functionally meaningful front and top directions of whole objects and target parts. These directions are cached and used to instantiate the functional alignment and semantic patch objectives. The optimization pipeline is implemented using PyTorch and nvdiffrast~\cite{laine2020modular}, with differentiable rasterization performed at a resolution of $64\times64$ pixels.

\subsection{Camera Configuration}
The camera anchor $\mathbf{c}_{a_t}$ is selected according to the action-conditioned target type and remains fixed throughout the corresponding action. For object--part actions, such as pressing, rotating, opening, and closing, the anchor is placed at the center of the manipulated part. For relational actions, such as picking, moving, and placing, the anchor is placed at the center of the parent object to preserve the surrounding interaction context. The manipulated object and, when applicable, the destination entity remain separate visual targets. The camera is oriented toward $\mathbf{c}_{a_t}$ together with the optimized in-plane roll. Independent camera panning and look-at offsets are not optimized in the current implementation.

\subsection{Differentiable Optimization}

We use multi-start initialization to reduce sensitivity to local optima. By default, 15 initial camera positions are distributed over the admissible azimuth and elevation domain. All initial candidates are evaluated, and the three candidates with the lowest objective losses are refined using Adam. The default learning rate is 0.08, and each candidate is optimized for at most 80 iterations. Optimization terminates early when the best objective value does not improve by more than $10^{-6}$ for five consecutive iterations. For motion-aware optimization, we sample three states along the articulated motion, including the start and end states, and aggregate their metric values when optimizing the viewpoint for the complete action motion.

\subsection{Objective Configuration}
The five objective weights for visible coverage, patch entropy, context-aware focus, functional alignment, and temporal continuity are set to 1.0 by default. A metric can be disabled by setting its weight to zero. The default visible-coverage target is 0.2, which provides an empirical balance between target legibility and preservation of the surrounding context.

For context-aware focus, we distinguish three types of entities: the manipulated target, the parent object that provides global structural context, and other parts of the model that provide local task context. By default, the task-importance distribution is derived from the relative visible projected areas of these entities under the functional front view determined by the semantic directional frame. This distribution preserves the relative visual prominence of the task-relevant entities under a semantically meaningful frontal composition. Users may override these default values to emphasize particular targets or contextual entities. The objective weights, entity-importance values, target visible coverage, number of initial candidates, number of refined candidates, learning rate, iteration limit, and number of motion samples are exposed through the graphical interface and configuration files. VISTA can therefore be adapted to different objects, actions, and visualization requirements without modifying the optimization code. By tuning these parameters, users can obtain higher-quality views for specific tasks.

\subsection{Runtime}

All visualization results are recorded on a machine equipped with an NVIDIA RTX 4090 GPU. The reported wall-clock optimization time for a single action includes multi-start candidate evaluation and Adam refinement, while offline LOD construction and cached VLM semantic-face inference are excluded. Optimization requires approximately 5--10 seconds per action for RealAppliance and approximately 20 seconds for the Chevrolet C8 and Forklift model. The resulting long-horizon animations are provided in the supplementary video.
}

\begin{figure*}[ht]
    \vspace{-2mm}
    \centering
    \includegraphics[width=0.9\linewidth]{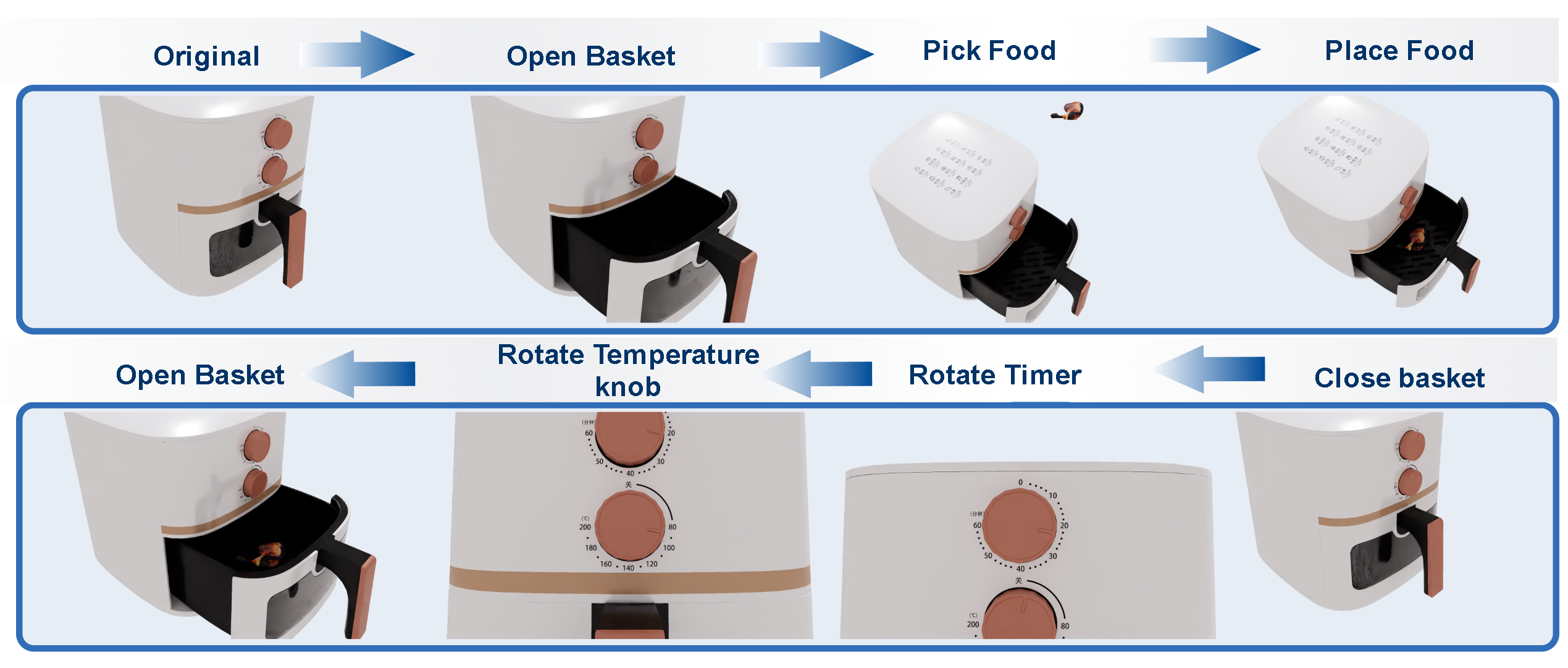}
    \vspace{-2mm}
    \caption{
\textbf{Use Case of VISTA on a long-horizon cooking task.}
Given an action sequence, VISTA selects a sequence of viewpoints that adapts to each manipulation step.
For coarse interactions (e.g., opening and closing the basket), the camera maintains a global view of the appliance, while for fine-grained operations (e.g., rotating knobs), it zooms in to reveal precise state information.
By aligning viewpoints with task-relevant targets at each step, VISTA preserves both the global task context and local detail, enabling clear and coherent visualization of the full task execution.
}
    \label{fig:case study}
    \vspace{-3mm}
\end{figure*}

\section{Use Case}
We demonstrate VISTA in~\autoref{fig:case study} through a long-horizon cooking task with an air fryer. The input to our system is an action sequence:

\textit{open(basket); pick(chicken); move(chicken, basket); place(chicken, basket); close(basket); rotate(timer knob, 90); rotate(temperature knob, 90); open(basket); pick(chicken)}.
For each action, VISTA automatically identifies the task-relevant target and optimizes a viewpoint that best explains the ongoing manipulation.

For \textit{open(basket)}, the action belongs to the first interaction type, object-part pair $(o,p)$, where the target is a single manipulated part. Here, the basket is defined as the target part, and the whole object is the air fryer. The selected viewpoint should therefore make the basket clearly visible and provide the essential context around the basket.

In contrast, \textit{move(chicken, basket)} belongs to the second interaction type, namely the relational tuple $(o_1,p,o_2)$, where the task depends on the spatial relationship between two entities. In this case, the chicken and the basket are jointly treated as target meshes, while the whole object remains the air fryer. These two targets are explicitly incorporated into different viewpoint quality metrics. The selected viewpoint should therefore simultaneously present the food item, the basket, and their relation, so that the moving process is easy to interpret.

\autoref{fig:visualization} illustrates one rendered image together with the visualizations of three metrics for that image. In this example, the target part is the timer knob. Therefore, in \autoref{fig:visible}, both target pixels and parent pixels are shown, as this metric measures the ratio of visible target pixels in the full image; the orange pixels indicate the target region. In \autoref{fig:part}, only the target is rendered, since parent occlusion is not considered in the patch entropy computation. Three semantic patches are shown in the patch entropy visualization, where the partition is determined by the semantic information. \autoref{fig:important} illustrates the visualization of the context-aware focus metric. This metric considers pixels from all parts as well as the parent pixels, and is computed as the Jensen--Shannon divergence between the observed pixel distribution and the desired context-focus distribution.

\section{Evaluation}
\highlight{We evaluate VISTA in four stages. First, we assess the reliability of the VLM used to extract the semantic front and up directions required by patch entropy and functional alignment. Second, we conduct a qualitative leave-one-objective-out ablation to illustrate the complementary effects of the individual viewpoint-quality metrics. Third, we perform a user study to evaluate the perceptual quality of the optimized camera trajectories in terms of action visibility, spatial context, camera-motion smoothness, and overall preference. Finally, we introduce a VLM-based perceptual benchmark to measure how different camera strategies affect appliance-state recognition, goal reasoning, and closed-loop control. Together, these evaluations examine the reliability of the semantic inputs, the contributions of the objective terms, and the effectiveness of VISTA for both human-facing task visualization and agent-facing visual observation.} 

\subsection{VLM Semantic Direction Extraction Evaluation}
\highlight{
To evaluate the reliability of the semantic information extracted by the VLM, we manually annotated the front and top directions of the whole object and every functional part in ten RealAppliance models, resulting in 68 instances comprising 10 whole objects and 58 parts. Each whole object was represented by six axis-aligned views, while each part was provided with six isolated views and six contextual views showing its location within the appliance. Using identical inputs and a fixed structured-output prompt, GPT-5.4-2026-03-05 achieved 92.65\% front accuracy, 95.59\% top accuracy, and 92.65\% joint accuracy, where joint accuracy requires both directions to be correct for the same instance. GPT-4o-2024-08-06 achieved 55.88\%, 91.18\%, and 55.88\%, respectively. These results demonstrate that an advanced VLM can reliably extract the semantic and reference-direction information required by our camera optimization framework, while GPT-4o’s lower front-direction accuracy indicates that model selection remains important.
}
\begin{figure*}[t]
    \centering
    \captionsetup[subfigure]{font=footnotesize,skip=2pt}

    \begin{subfigure}[t]{0.315\textwidth}
        \vspace{0pt}
        \centering
        \includegraphics[width=\linewidth]{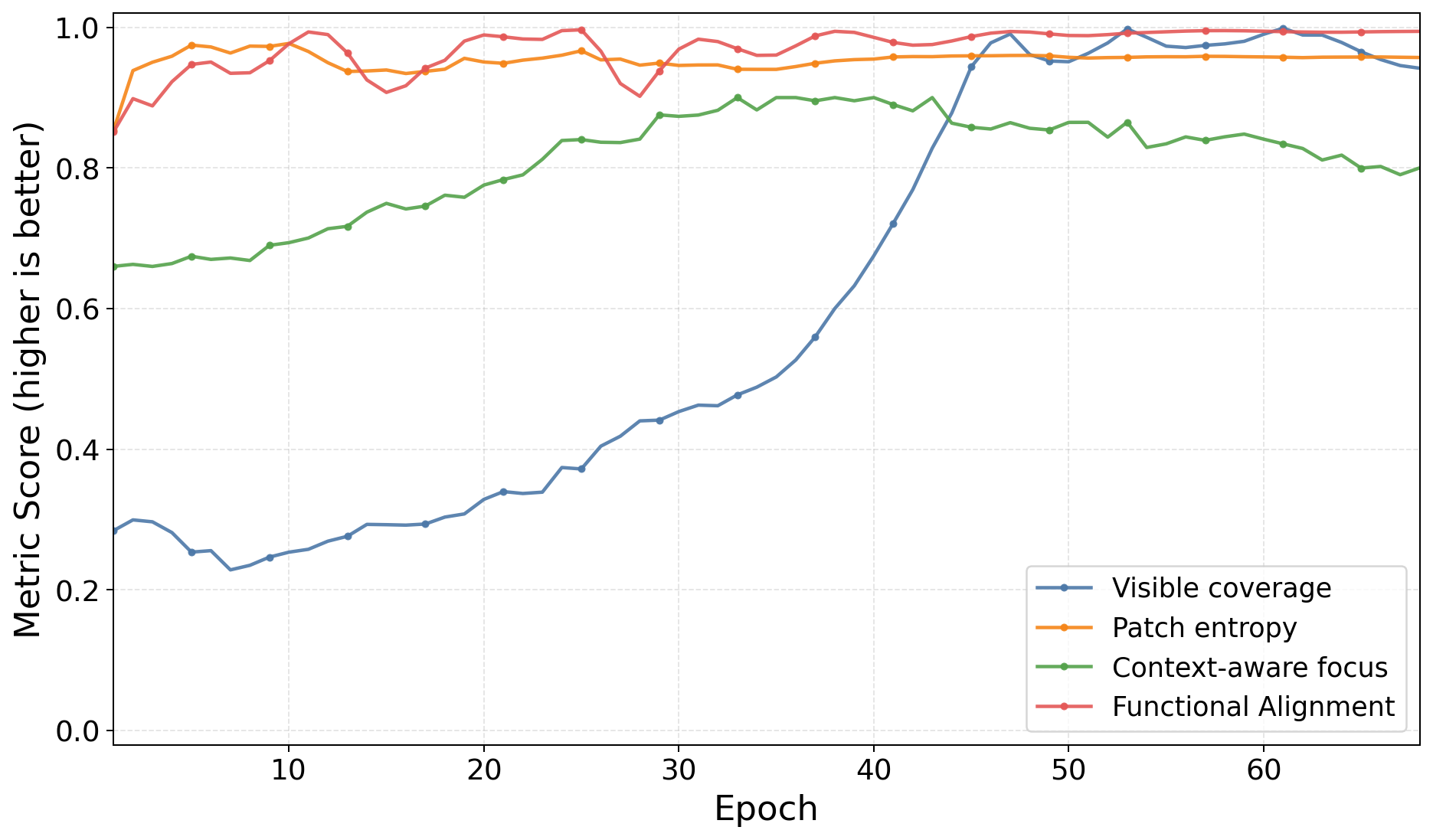}
        \caption{Metric scores}
        \label{fig:metric_scores}
    \end{subfigure}
    \hfill
    \begin{subfigure}[t]{0.315\textwidth}
        \vspace{0pt}
        \centering
        \includegraphics[width=\linewidth]{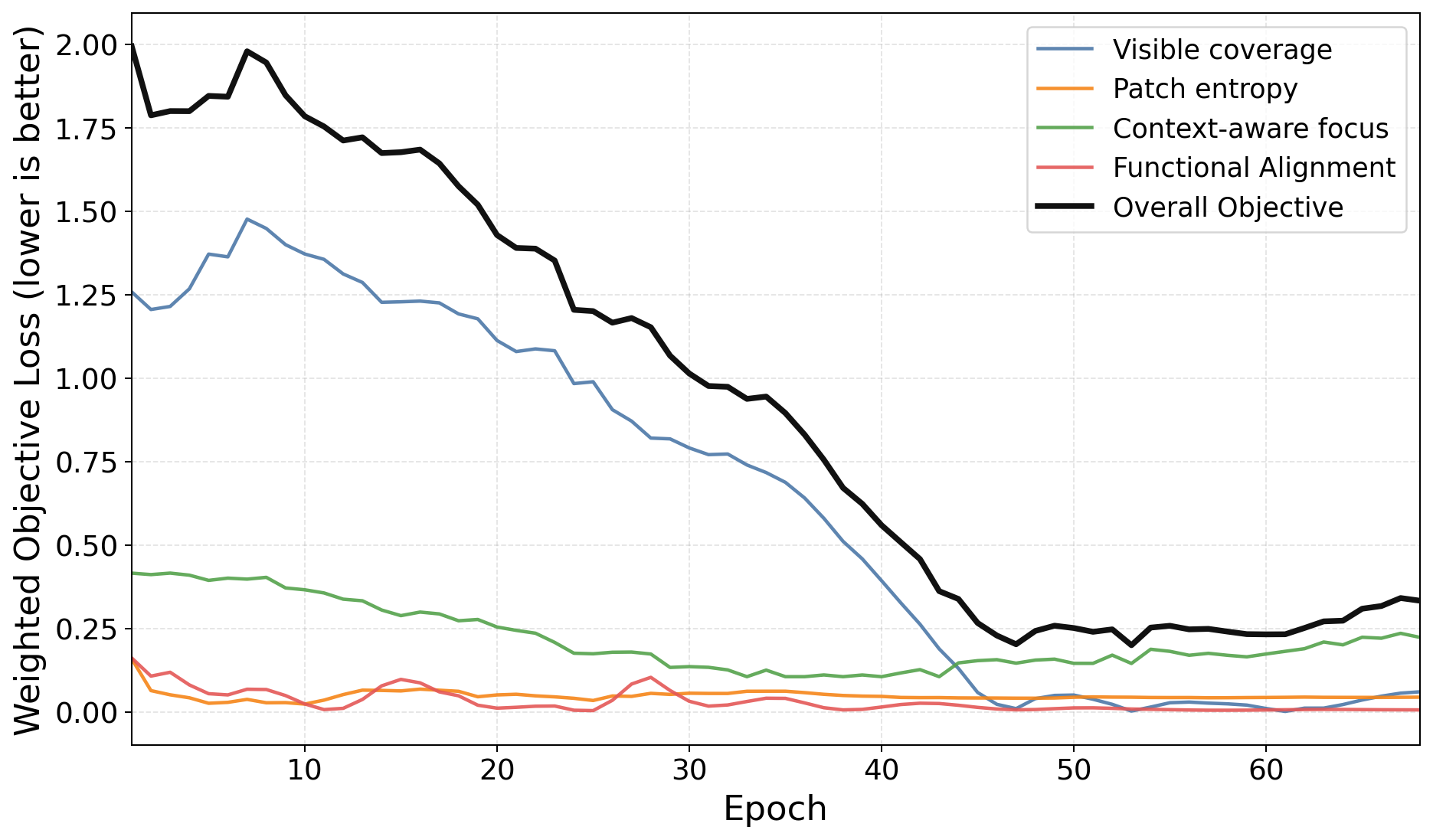}
        \caption{Objective losses}
        \label{fig:objective_losses}
    \end{subfigure}
    \hfill
    \begin{subfigure}[t]{0.315\textwidth}
        \vspace{0pt}
        \centering
        \includegraphics[width=\linewidth]{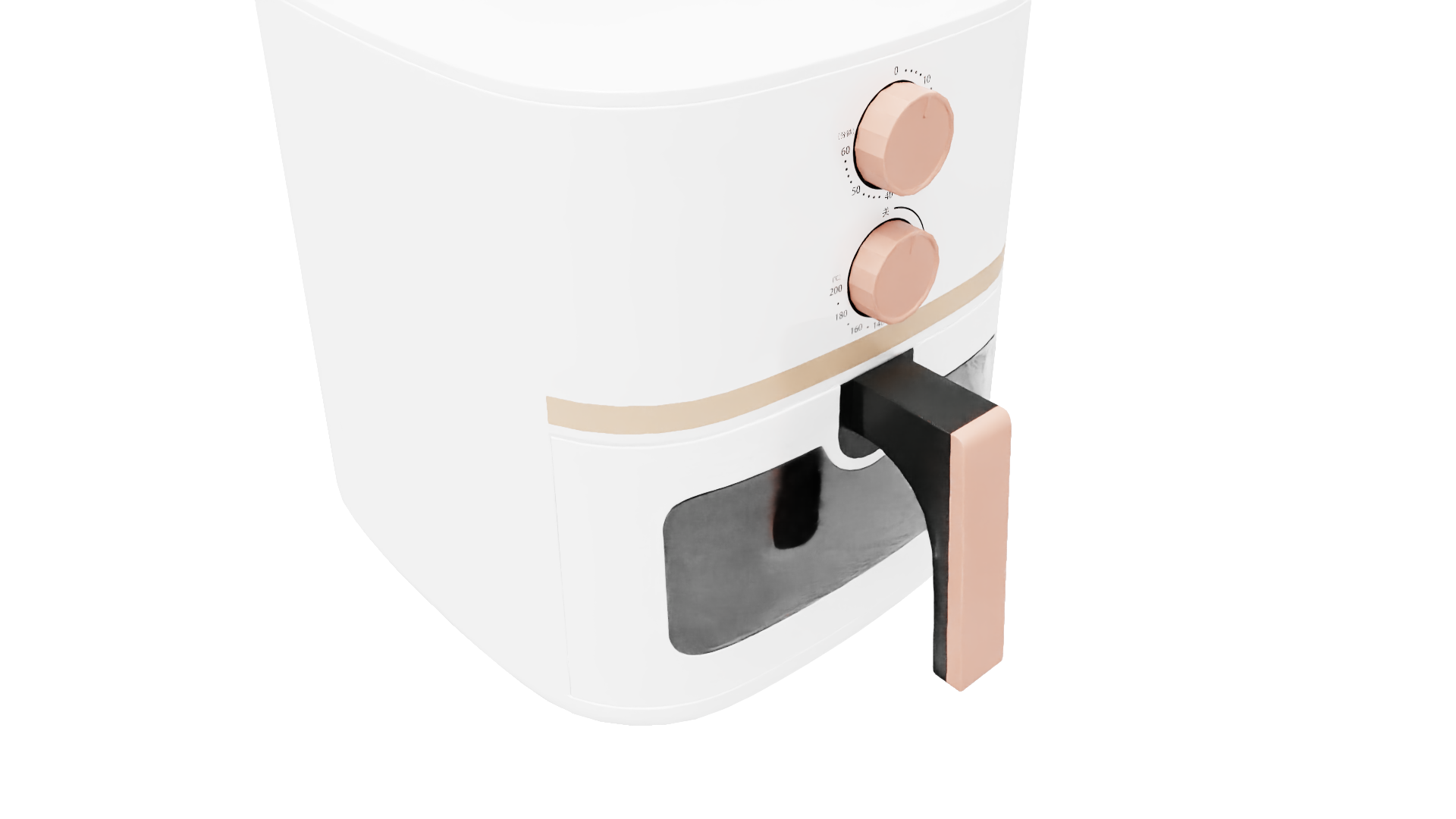}
        \caption{Full objective}
        \label{fig:full_objective}
    \end{subfigure}

    \par\vspace{0.6em}

    \begin{subfigure}[t]{0.235\textwidth}
        \vspace{0pt}
        \centering
        \includegraphics[width=\linewidth]{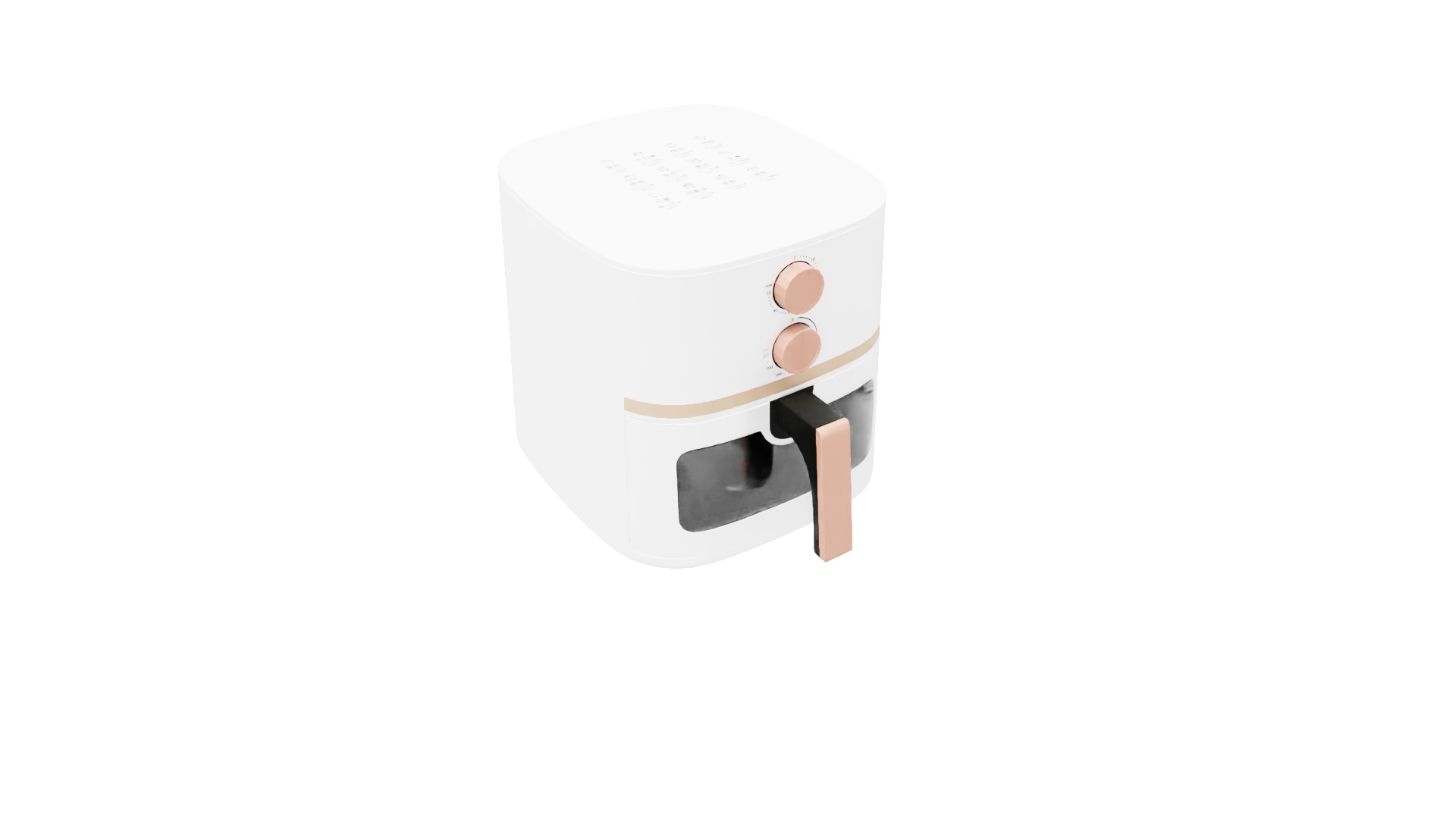}
        \caption{w/o Visible coverage}
        \label{fig:wo_coverage}
    \end{subfigure}
    \hfill
    \begin{subfigure}[t]{0.235\textwidth}
        \vspace{0pt}
        \centering
        \includegraphics[width=\linewidth]{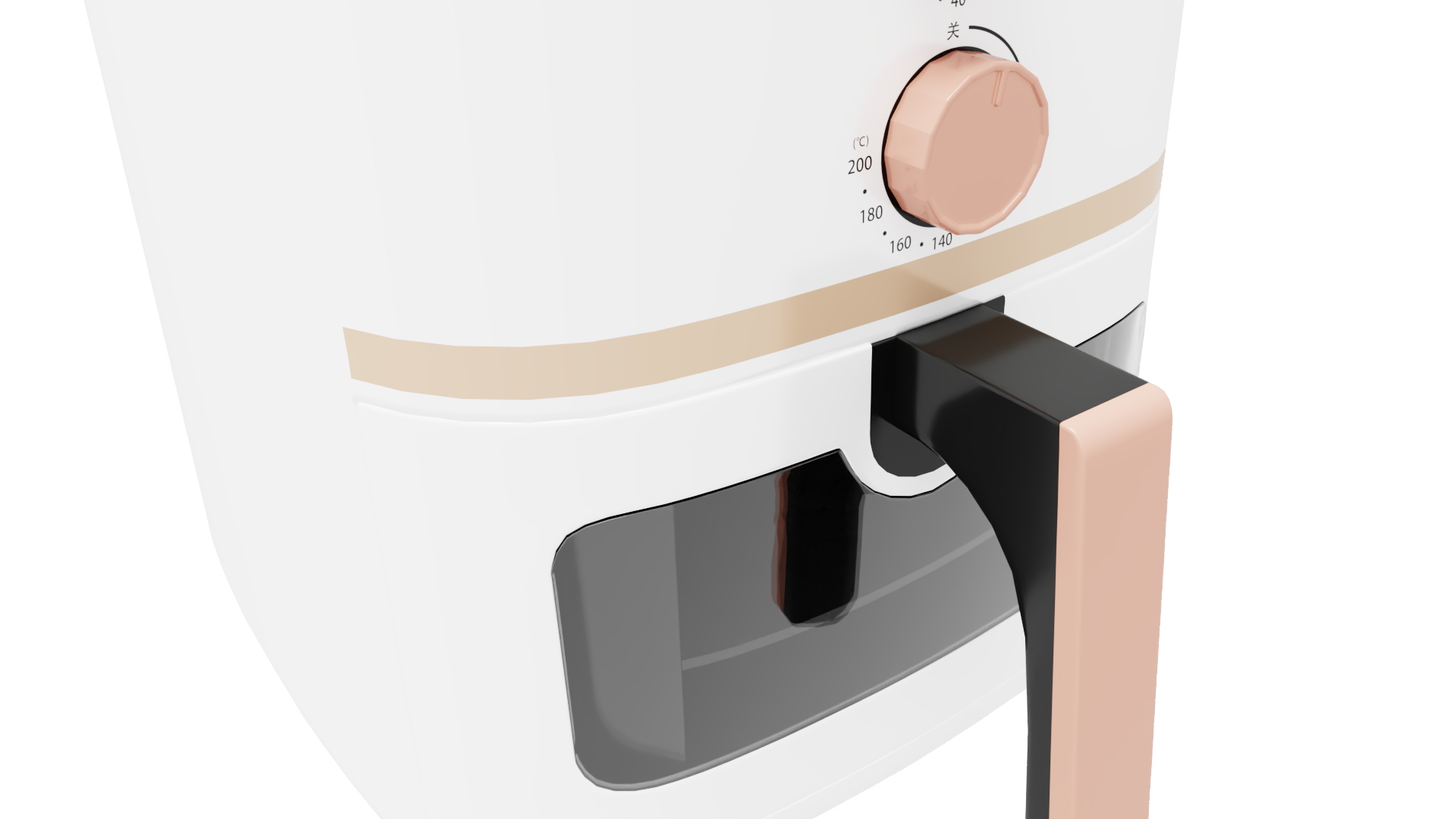}
        \caption{w/o Context-aware focus}
        \label{fig:wo_context}
    \end{subfigure}
    \hfill
    \begin{subfigure}[t]{0.235\textwidth}
        \vspace{0pt}
        \centering
        \includegraphics[width=\linewidth]{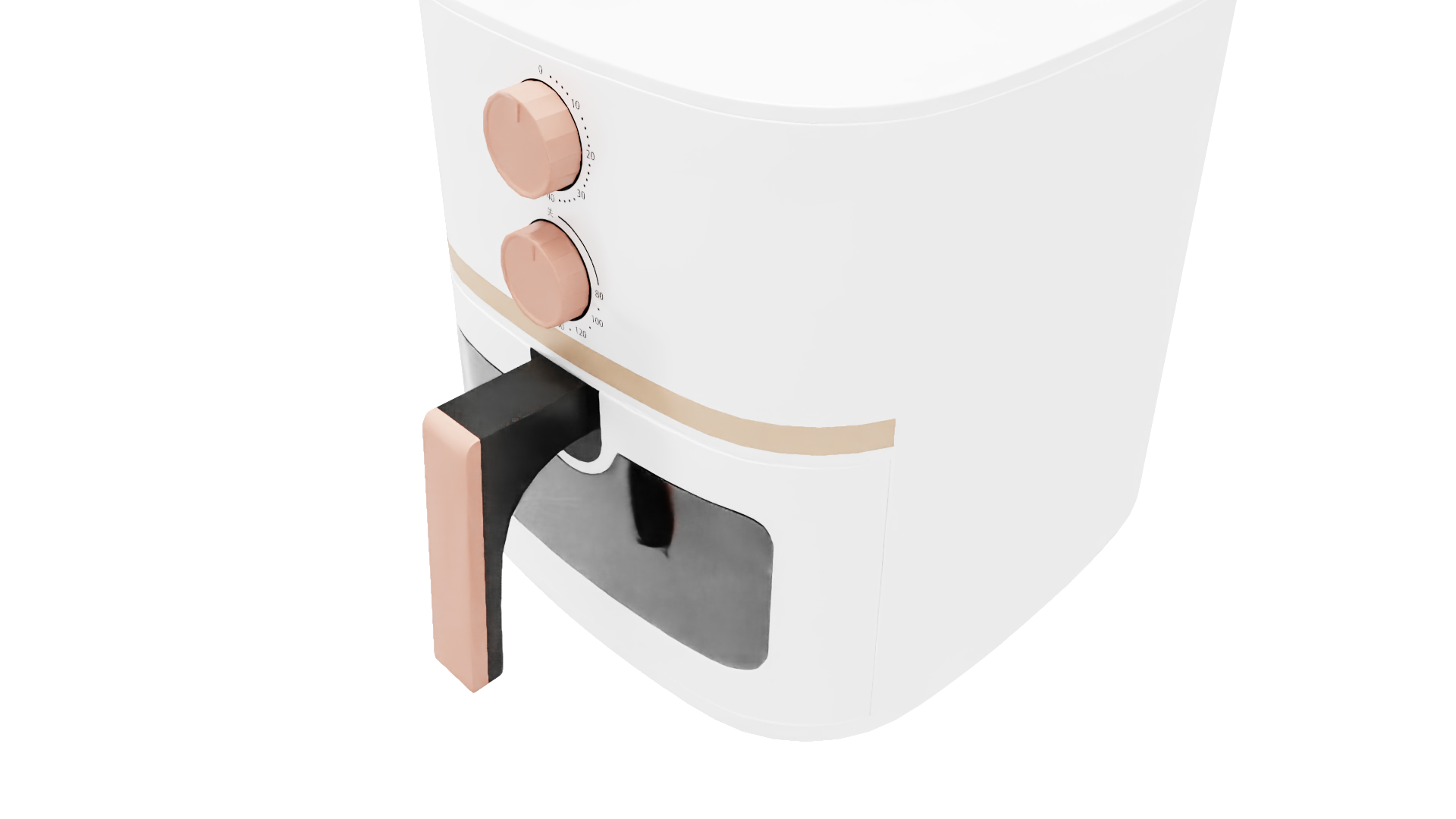}
        \caption{w/o Patch entropy}
        \label{fig:wo_patch}
    \end{subfigure}
    \hfill
    \begin{subfigure}[t]{0.235\textwidth}
        \vspace{0pt}
        \centering
        \includegraphics[width=\linewidth]{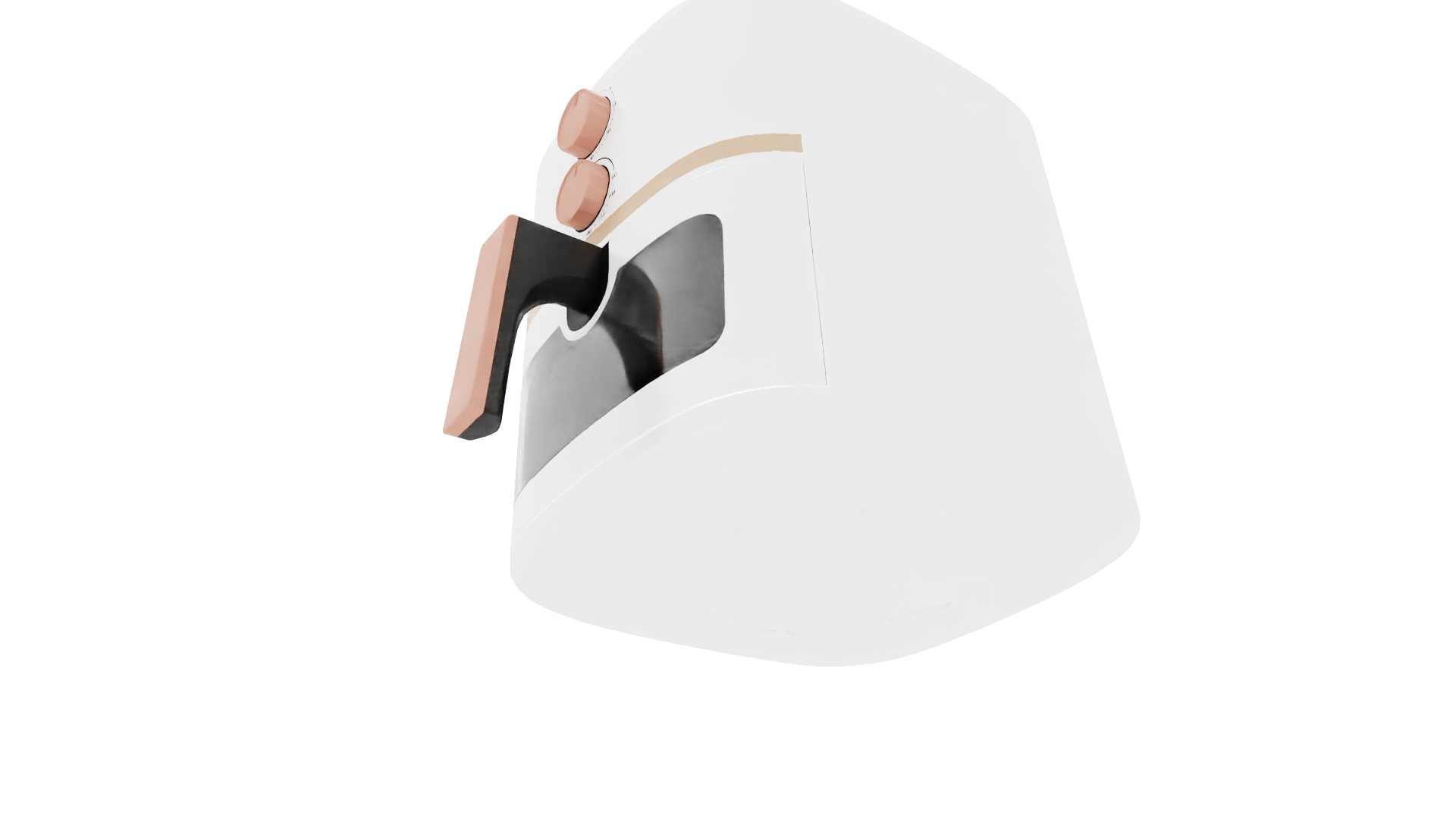}
        \caption{w/o Functional Alignment}
        \label{fig:wo_function}
    \end{subfigure}

    \caption{
    \highlight{
    Optimization curves and qualitative objective ablations for the open(basket) action.
    First row shows the metric scores, objective losses, and the
    optimized view using the full objective. The second row shows the
    resulting views when each metric is removed.}
    }
    \label{fig:objective_ablation}
\end{figure*}
\subsection{Qualitative Objective Ablation}
\highlight{
We conduct a qualitative leave-one-objective-out ablation to examine how the individual per-view metrics influence the optimized camera configuration. As shown in \autoref{fig:objective_ablation}, we optimize the viewpoint for the action \textit{open(basket)} using the complete objective and four variants in which visible coverage, context-aware focus, patch entropy, or functional alignment is removed individually. Temporal continuity is not included in this single-view comparison because it is defined between consecutive camera configurations and therefore requires a sequence-level evaluation.

The optimization curves in \autoref{fig:objective_ablation} show that the individual metric scores improve while the weighted objective loss decreases during optimization. With the complete objective, VISTA produces a close and functionally aligned view that clearly presents the manipulated basket while retaining sufficient context from the surrounding appliance.

Removing visible coverage results in a substantially more distant viewpoint, causing the manipulated basket to occupy only a small region of the image. This demonstrates the role of visible coverage in maintaining sufficient target prominence. Removing context-aware focus changes the relative image-space allocation between the basket and its parent appliance, producing a composition that does not preserve the same balance between local action detail and surrounding context. For example, as shown in the figure, removing context-aware focus causes the top knob, a task-relevant contextual part, to disappear from the resulting view.

Without patch entropy, the optimizer produces a more oblique view with a less balanced presentation of the target’s directional surface regions. In contrast, removing functional alignment results in a viewpoint from obliquely below the object, without respecting its semantically meaningful front and up directions. Although patch entropy and functional alignment use the same semantic directional frame, these results illustrate their complementary roles: patch entropy encourages the exposure of diverse target surfaces, whereas functional alignment prevents semantically inappropriate viewing directions and favors an operationally meaningful composition.

Overall, the complete objective provides the most balanced composition among target visibility, structural informativeness, functional orientation, and contextual presentation. This experiment is intended as a qualitative illustration of the behavior of the individual objective terms rather than a statistical analysis of their relative importance or weight sensitivity. A more comprehensive quantitative evaluation across actions, object categories, and objective-weight configurations remains an important direction for future work.}
\subsection{User Study}
\label{sec:user_study}
\begin{figure}[t]
    \centering
    \includegraphics[width=\columnwidth]{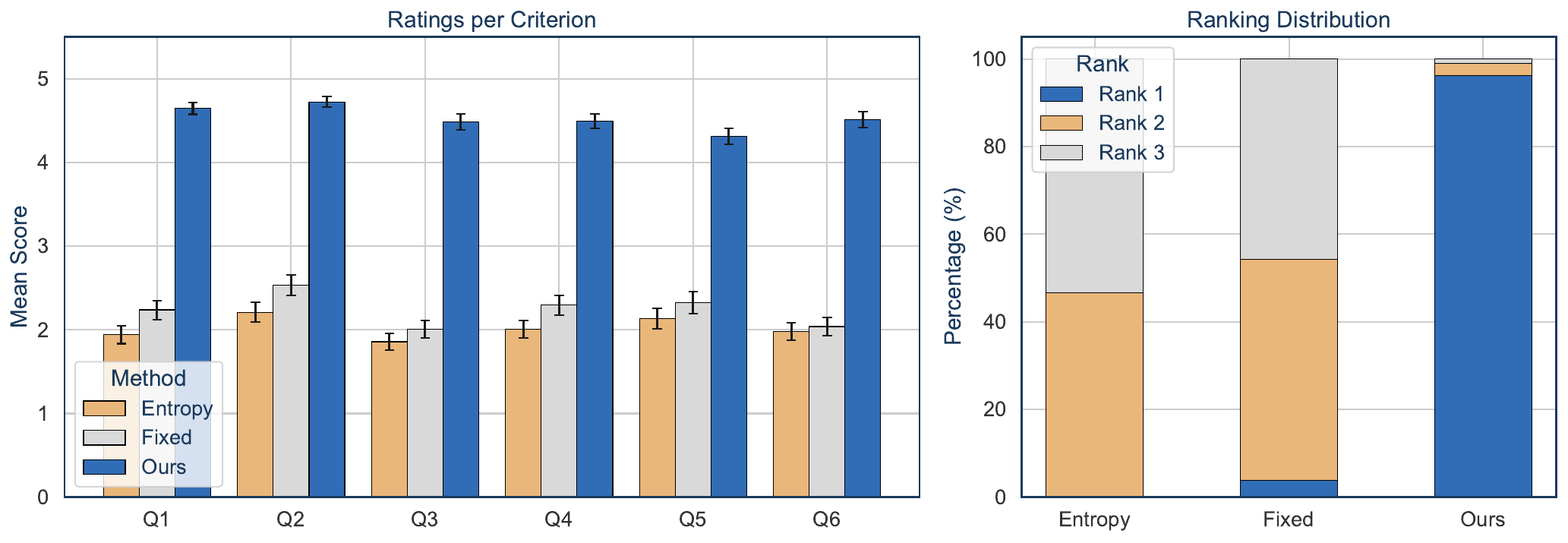}
    \vspace{-2mm}
    \caption{Mean ratings per criterion for each method (left). Ranking distribution across all tasks(right).}
    \label{fig:user_study}
    \vspace{-2mm}
\end{figure}
We conducted a user study with 21 participants to evaluate the perceptual quality on the RealAppliance datasets.  The camera trajectories are optimized by our method against two baselines: a fixed camera placement (\textit{Fixed}) and a camera position selected using viewpoint entropy~\cite{Vazquez2001ViewpointEntropy} (\textit{Entropy}). The fixed camera position is obtained from the default position in the RealAppliance dataset. To obtain the viewpoint with the highest viewpoint entropy score, we sample candidate viewpoints on a sphere with a diameter of 1.5 times the object bounding box diagonal and select the one with the maximum entropy value. 

\highlight{We did not include the learned human-preference baseline in the user study because, similar to viewpoint entropy, it produces a static whole-object view and does not adapt to changes in the action target. Because participants were required to watch, rate, and rank every camera trajectory, including both static optimized-view baselines would have increased the study duration and participant fatigue without introducing a substantially different camera strategy. We therefore retained entropy as a deterministic and interpretable representative of static optimized viewpoint selection.} Our experiment was approved by our university’s IRB. At the start of the experiment, we obtained informed consent from all participants.

We employed a within-subjects design, where every participant evaluated all three methods across all tasks, enabling direct comparison and reducing individual variability.
Participants viewed camera-generated videos of five appliance interaction sequences: (1)~airfryer, cook cycle, (2)~bread machine, bake bread, (3)~airfryer, air fry session, (4)~oven, adjust controls, and (5)~bread machine, bake cycle. Each task depicted the same interaction captured by all three camera methods.
For each task, participants watched three videos (one per method) presented in a randomized order. They then rated each method on six criteria using a 5-point Likert scale (1~=~worst, 5~=~best) and ranked the three methods from best to worst. The presentation order of methods was randomized per task per participant across all six possible permutations to mitigate ordering bias.

Participants rated each method on the following six criteria:
\begin{itemize}
    \item Q1: How well the action is visible in the video.
    \item Q2: How well the camera provides spatial context.
    \item Q3: How natural and smooth the camera motion feels.
    \item Q4: How well the camera avoids occluding.
    \item Q5: General quality of the camera trajectory.
    \item Q6: Subjective preference for the method.
\end{itemize}

As shown in Figure~\ref{fig:user_study}, our method received significantly higher ratings than both baselines across all six evaluation criteria. Our method achieved a mean score of 4.53 (std = 0.88), compared to 2.24 (std = 1.20) for Fixed and 2.02 (std = 1.13) for Entropy. Participants ranked our method first in 96\% of all tasks (101 out of 105), with a mean rank of 1.05 (1 = best). Fixed and Entropy received mean ranks of 2.42 and 2.53, respectively. (Figure~\ref{fig:user_study}).

\begin{table*}[t]
\centering
\small
\setlength{\tabcolsep}{3.5pt}
\begin{tabular}{l cccc cccc cccc cccc}
\toprule
& \multicolumn{4}{c}{\textbf{Fixed}}
& \multicolumn{4}{c}{\textbf{Entropy}}
& \multicolumn{4}{c}{\textbf{Learning}}
& \multicolumn{4}{c}{\textbf{Adaptive (Ours)}} \\
\cmidrule(lr){2-5}\cmidrule(lr){6-9}\cmidrule(lr){10-13}\cmidrule(lr){14-17}
\textbf{Model}
  & T1$\uparrow$ & T2$\uparrow$ & T3-SR$\uparrow$ & T3-Err$\downarrow$
  & T1$\uparrow$ & T2$\uparrow$ & T3-SR$\uparrow$ & T3-Err$\downarrow$
  & T1$\uparrow$ & T2$\uparrow$ & T3-SR$\uparrow$ & T3-Err$\downarrow$
  & T1$\uparrow$ & T2$\uparrow$ & T3-SR$\uparrow$ & T3-Err$\downarrow$ \\
\midrule
GPT-5.4
  & 60.0 & 70.0 & 45.8 & 1.03
  & 52.0 & 58.0 & 62.5 & 1.00
  & 56.0 & 54.0 & 62.5 & 0.85
  & \textbf{88.0} & \textbf{86.0} & \textbf{81.9} & \textbf{0.36} \\
GPT-4o
  & 52.0 & 44.0 & 15.3 & 1.79
  & 44.0 & 34.0 &  0.0 & 2.53
  & \textbf{56.0} & 46.0 & 15.3 & 1.97
  & \textbf{56.0} & \textbf{68.0} & \textbf{55.6} & \textbf{0.90} \\
Gemini-2.5-Flash
  & 40.0 & 50.0 & 30.6 & 1.97
  & 48.0 & 40.0 & 20.8 & 2.19
  & 52.0 & 64.0 & 13.9 & 2.19
  & \textbf{64.0} & \textbf{84.0} & \textbf{59.7} & \textbf{0.96} \\
\midrule
\textit{Avg.}
  & 50.7 & 54.7 & 30.6 & 1.60
  & 48.0 & 44.0 & 27.8 & 1.91
  & 54.7 & 54.7 & 30.6 & 1.67
  & \textbf{69.3} & \textbf{79.3} & \textbf{65.7} & \textbf{0.74} \\
\bottomrule
\end{tabular}
\caption{
  VLM perceptual benchmark results across three tasks and four camera configurations.
  \textbf{T1}: state recognition accuracy (\%, Task~1).
  \textbf{T2}: goal-decision accuracy.
  \textbf{SR}: closed-loop control success rate (\%, Task~3).
  \textbf{Err}: mean StateErr $=\mathbb{E}[|\text{idx}(g)-\text{idx}(s_T)|]$ in discrete steps (Task~3).
}
\label{tab:vlm_results}
\end{table*}

\subsection{VLM Perceptual Benchmark}
Prior closed-loop benchmarks for embodied appliance manipulation~\cite{gao2025realappliancelethighfidelityappliance} evaluate whether a VLM can predict the correct next action given a rendered observation, task instruction, the full task history, and product manuals. While this protocol evaluates end-to-end task performance, it inherently conflates \emph{perceptual quality} with \emph{task reasoning}. A model’s failure may stem either from ambiguous or insufficient visual observations or from a lack of procedural knowledge required to correctly sequence actions. Consequently, such benchmarks are not well suited for evaluating viewpoint quality, because their outcomes do not depend solely on the visual information provided by the camera, but also on the model’s ability to understand the task and reason about the required action. Moreover, given the overall complexity of these benchmarks relative to current VLM capabilities, models tend to rely heavily on semantic priors for decision-making. This tendency is further evidenced by our ablation study (see Supplementary Material Section 3), where models exhibit non-negligible performance even when provided with a completely black image.

We therefore propose a new benchmark that isolates perceptual difficulty from task reasoning to measure the different viewpoints. The benchmark decouples appliance-state perception into three progressively complex tasks, each requiring only a single rendered image and a minimal contextual prompt, thereby reducing the influence of additional semantic information and encouraging the evaluation to depend primarily on the visual evidence provided by the viewpoint. This design makes performance differences between camera configurations directly attributable to viewpoint quality, not confounding factors.

We formulate all tasks under a unified observation-state-action framework. Each appliance part is associated with a discrete state space $S$, defined by an ordered vocabulary (e.g., knob positions or binary states), and a rendering function $R: S \rightarrow O$ that maps states to RGB observations. 
Given an observation $o \in O$, the VLM is required to infer, compare, or control the underlying state.

\noindent\textbf{Task 1: State Recognition.}
This task evaluates direct perception as a mapping
$ f: O \rightarrow S$,
where the model predicts the current state $\hat{s} = f(o)$.
The metric is classification accuracy:
\begin{equation}
    \text{Acc}_{\text{T1}} = E_{(o, s^*)} \left[\mathbf{1}(\hat{s} = s^*)\right].
\end{equation}
We denote the ground-truth state as $s^* \in S$ and the predicted state as $\hat{s} \in S$.  For example, consider an air fryer whose timer knob has vocabulary ${S} = [\texttt{off},\ \texttt{10min},\ \ldots,\ \texttt{60min}]$.
In a Task~1 example, the knob could be set to \texttt{30min} and the VLM must read the label from a single image; a suboptimal viewpoint that blurs the dial markings could cause misclassification, e.g. \texttt{"20min"}.

\noindent\textbf{Task 2: Goal Decision.}
Given an observation $o \in O$ and a goal state $g \in S$, the model performs a binary judgment of whether the part has reached the specified goal $f: O \times S \rightarrow \{0,1\}$ producing prediction $\hat{y} = f(o, g)$.
The ground-truth label is defined as $y^* = \mathbf{1}(s^* = g)$, where $s^* \in \mathbb{S}$ denotes the underlying state of the observation.
For each (model, part, current-state) triple, one positive sample (goal $= s^*$) and one negative sample (goal selected as the state with the largest index distance from $s^*$) are generated, yielding a perfectly balanced 1:1 dataset.
The metric is classification accuracy:
\begin{equation}
    \text{Acc}_{\text{T2}} = E_{(o,g,y^*)}\left[\mathbf{1}(\hat{y} = y^*)\right].
\end{equation}
In Task~2, one image is paired with the goal \texttt{"60min"} and the model must answer state \texttt{"60min"} and \texttt{YES}; a model that cannot predict the correct state and reason whether the goal is achieved if it can not resolve the precise dial position.

\noindent\textbf{Task 3: Closed-Loop Delta Control.}
Closed-loop control is an iterative process in which the model observes the current state, takes an action, and receives updated observations reflecting its effect. 
Unlike single-step tasks, it requires consistent perception and action sequencing over time, with errors accumulating across steps. 

We model control as a discrete sequential decision process over $S$.
At each step $t$, the model observes $o_t = R(s_t)$ and predicts an action $a_t \in A = \{\texttt{increase}, \texttt{decrease}, \texttt{stop}\}$.
The environment transitions deterministically $s_{t+1} = T(s_t, a_t)$,
where $T$ shifts the state by one discrete step or terminates if \texttt{stop} is issued.
An episode succeeds if the model reaches the goal state $g$ and outputs \texttt{stop} within $T_{\max}=10$ steps.
We report the success rate and the final state error:
\begin{equation}
    \text{SR} = \frac{N_{\text{succ}}}{N_{\text{ep}}}, \quad
    \text{StateErr} = E\left[\bigl|\,\text{idx}(g) - \text{idx}(s_T)\,\bigr|\right],
\end{equation}
where $\text{idx}(\cdot)$ denotes the position of a state in the ordered vocabulary $S$ (e.g., $\text{idx}(\texttt{off})=0$, $\text{idx}(\texttt{10min})=1$, \ldots), and $s_T$ is the final state at episode termination -- either by a \texttt{stop} action or upon reaching $T_{\max}$.
StateErr thus measures residual distance to the goal in discrete steps, and equals zero for all successful episodes.
In Task~3, starting from \texttt{10min} with a goal \texttt{40min}, the adaptive camera enables the model to issue \texttt{increase} three times in succession and reach the goal in exactly three steps (StateErr $= 0$); under a fixed camera, the same model cannot reliably read the current position and oscillates between \texttt{off} and \texttt{10min} for all ten steps, terminating with StateErr $= 3$.

\textbf{Baselines.}
We evaluate four camera configurations. 
(1) fixed camera: a static viewpoint determined by appliance geometry, following RealAppliance~\cite{gao2025realappliancelethighfidelityappliance}. 
(2) human preference view: a viewpoint learned to approximate human-preferred camera placement~\cite{hartwig2022learning}. We utilize the second stage model from their paper, where the model could return a human preference score given a rendered image.  
(3) entropy view: a viewpoint optimized by maximizing viewpoint entropy, encouraging visually informative observations. 
(4) adaptive Camera (Ours): our task-aware camera optimization method. Both human preference view and viewpoint entropy view are calculated through samples on a sphere with a diameter of 1.5 times the object bounding box diagonal

\textbf{Result.}
Table~\ref{tab:vlm_results} summarizes performance across all tasks and camera configurations. Averaged over all VLMs, the adaptive camera consistently achieves the best results: 69.3\% on T1 (vs.\ 54.7\% Learning, 50.7\% Fixed, 48.0\% Entropy), 79.3\% on T2 (vs.\ 54.7\% Fixed/Learning, 44.0\% Entropy), and 65.7\% SR with 0.74 StateErr on T3 (vs.\ $\leq$30.6\% SR and $\geq$1.60 Err for all static baselines), corresponding to a \textbf{$2.1\times$} improvement in closed-loop success rate.
The gain is consistent across all models.
GPT-5.4 improves from 60.0\% to 88.0\% on T1 and from 45.8\% to 81.9\% SR on T3; 
Gemini-2.5-Flash improves from 50.0\% to 84.0\% on T2 and from 30.6\% to 59.7\% SR on T3; 
GPT-4o improves from 44.0\% to 68.0\% on T2 and recovers from 0\% SR under Entropy to 55.6\% under the adaptive camera.

Overall, these results demonstrate that adaptive camera planning plays a critical role in enabling robust performance under sequential control, which is important for embodied AI systems.

\section{Discussion}

\highlight{

\textbf{Objective alternatives and ablation.}
Our five objectives represent one operationalization of view quality rather than the only possible formulation. Alternative criteria differ in both their supervision requirements and optimization properties. Visible coverage can be expressed through image-space framing constraints~\cite{gleicher1992through} or downstream task-recognition confidence ~\cite{gualtieri2017viewpoint}. Structural informativeness can be defined using viewpoint or surface-information entropy~\cite{Vazquez2001ViewpointEntropy,page2003shape}. Functional alignment may instead be derived from joint or motion axes, interaction-surface normals, or affordance-specific preferred view regions. Context can be controlled through target-to-saliency alignment, or cinematographic composition rules\cite{he1996virtual}.

\textbf{Differentiable visibility and non-convex optimization} Soft pixel visibility provides useful gradients near silhouettes and occlusion boundaries, but it does not solve global visibility. If a target is fully hidden behind a large closed surface, small camera perturbations may not change its visibility, creating a flat or locally uninformative region even when a visible viewpoint exists elsewhere. The combined objective is also non-convex because visibility, context, semantic alignment, and continuity can favor different camera configurations. We mitigate this limitation using spatially distributed multi-start initialization followed by Adam refinement of the most promising candidates. This combines approximate global exploration with differentiable local optimization, but does not guarantee a global optimum: narrow visible regions may be missed, and no admissible camera can recover the target if it remains occluded throughout the action.

\textbf{Failure cases and limitations.}
Although VISTA conditions viewpoint planning on the structured action target and action type, it does not yet model richer action semantics such as motion direction, contact geometry, affordances, or causal state transitions.  For example, insertion, pouring, locking, and other state-dependent transformations may require views that reveal the interaction axis, the receiving region, or the before-and-after state rather than merely keeping the manipulated part visible. Incorporating action semantics into target definition, semantic direction estimation, and viewpoint optimization is therefore an important direction for future work.

The framework also depends on the quality of its geometric and semantic annotations. Incorrect or ambiguous front and top directions can cause functional alignment to favor an inappropriate side of the object, particularly for symmetric objects or parts without a clearly defined functional front. Similarly, inaccurate part segmentation, bounding boxes, or parent--target relationships can distort the coverage and contextual-focus objectives. Setting the visible-coverage target too high may produce an excessively zoomed-in view that removes useful surrounding context or clips the target during motion, whereas setting it too low may make subtle state changes difficult to observe. Sparse motion sampling can additionally miss short-lived occlusions or contacts occurring between sampled states. Patch entropy remains a lightweight geometric view-diversity prior rather than a complete measure of semantic informativeness. For a sphere or another isotropic and visually uniform object, directional patches may be geometrically distinct but perceptually indistinguishable. Thin structures, small functional features, and fine occlusion relationships may also be poorly represented by the simplified geometry or finite rasterization resolution. 

}

\section{Conclusion}

We presented VISTA, a differentiable viewpoint-planning framework for object-centered long-horizon task visualization. Rather than selecting a generic best view of an entire object, VISTA constructs action-conditioned object--part or relational targets and optimizes camera configurations to balance visible coverage, structural information, functional alignment, contextual focus, and temporal continuity. The framework supports both human-facing narrative visualization and agent-facing active observation.

\bibliographystyle{IEEEtran}
\bibliography{references}

\begin{IEEEbiography}[{\includegraphics[width=1in,height=1.35in]{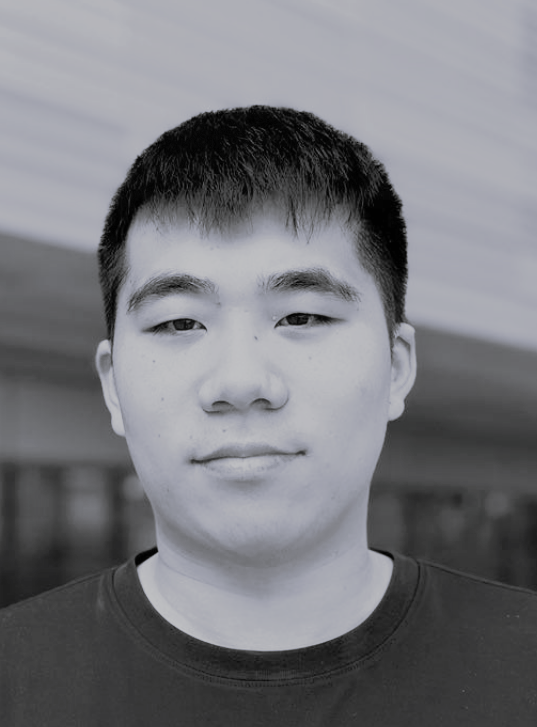}}]{Donggang Jia} is a Computer Science PhD student at King Abdullah University of Science and Technology (KAUST), where he is part of the nanovisualization group. He received a B. Eng degree in Computer Science from the Beijing University of Posts and Telecommunications, China, and an MSc in Artificial Intelligence from the University of Southampton, UK. His research interests include scientific visualization, human-computer interaction, and computer graphics.
\end{IEEEbiography}
\vspace{-10mm}
\begin{IEEEbiography}[{\includegraphics[width=1in,height=1.35in]{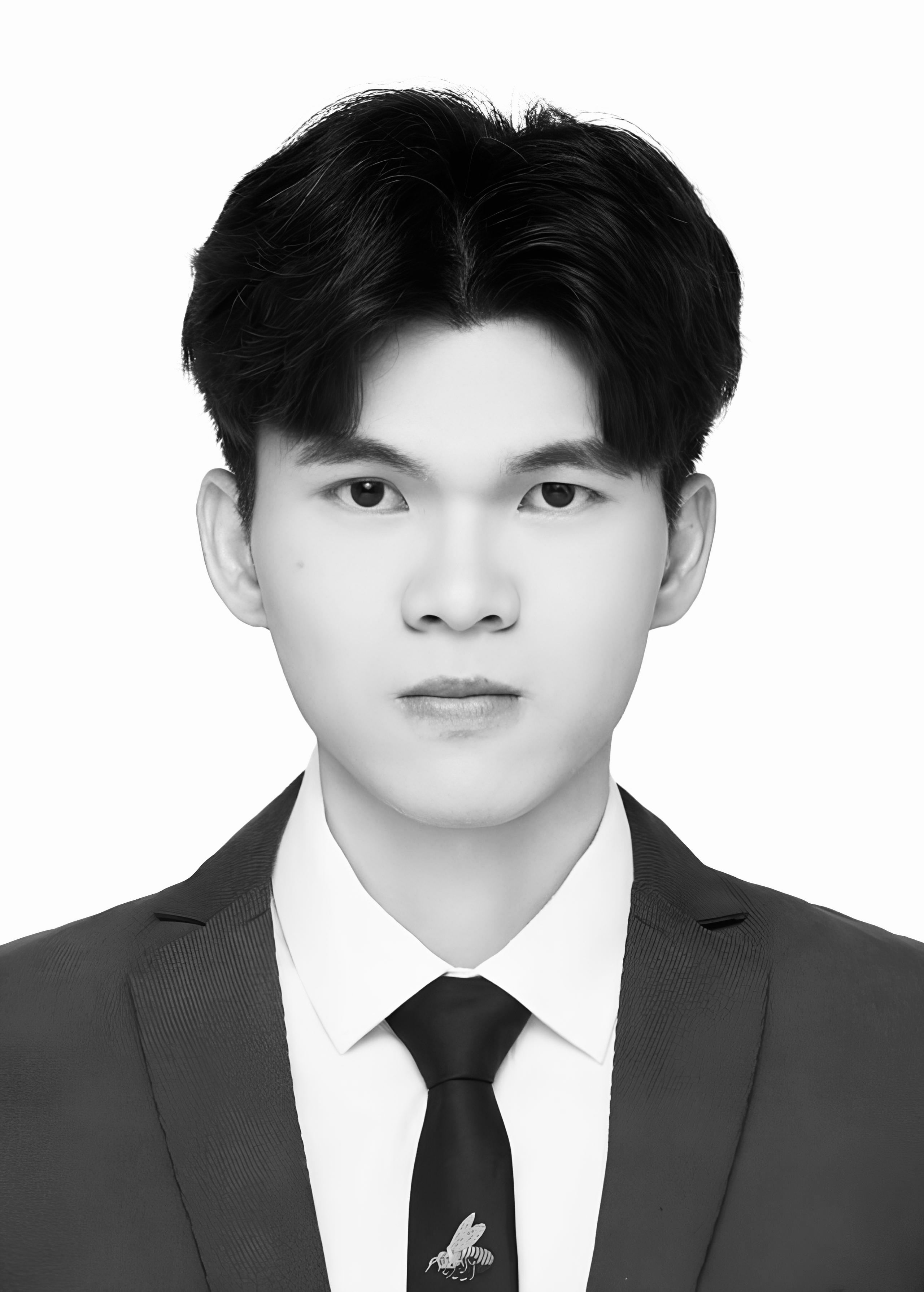}}]{Da Li} is a Computer Science PhD student at King Abdullah University of Science and Technology (KAUST) under the supervision of Prof. Ivan Viola. He received his B.Eng. degree from the University of Electronic Science and Technology of China (UESTC), China, and his M.Sc. degree in Computer Science from KAUST under the supervision of Prof. Ivan Viola. His research interests include 3D reconstruction, computer graphics, and 3D computer vision.
\end{IEEEbiography}
\vspace{-10mm}
\begin{IEEEbiography}[{
\includegraphics[width=1in]{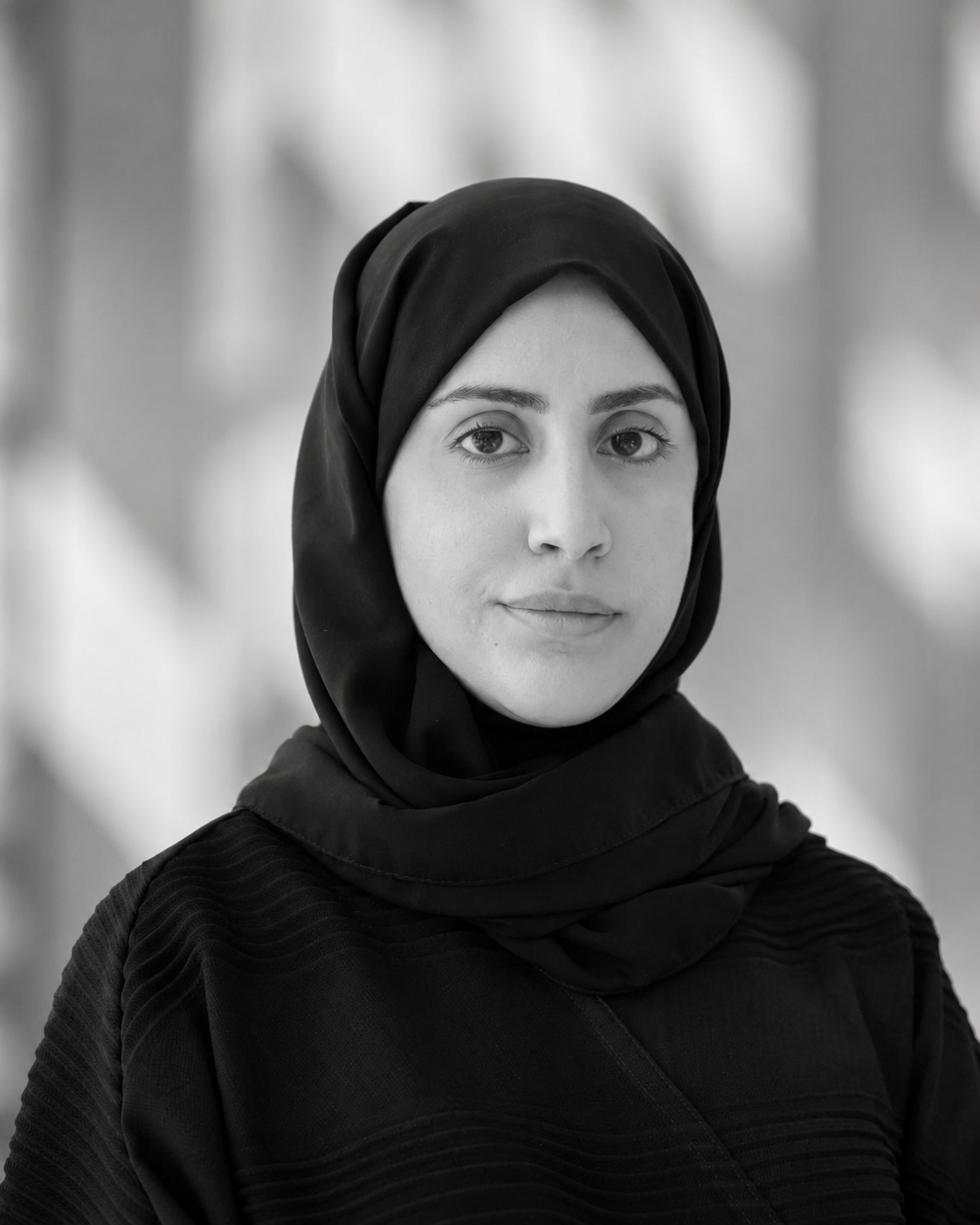}}]{Zainab Alsuwaykit} is currently a PhD student in Computer Science at King Abdullah University of Science and Technology (KAUST). She earned her master’s degree from Wichita State University, Kansas, USA, in 2018. Her research interest is graph drawing and visualization.
\end{IEEEbiography}
\vspace{-10mm}
\begin{IEEEbiography}[{\includegraphics[width=1in,height=1.35in]{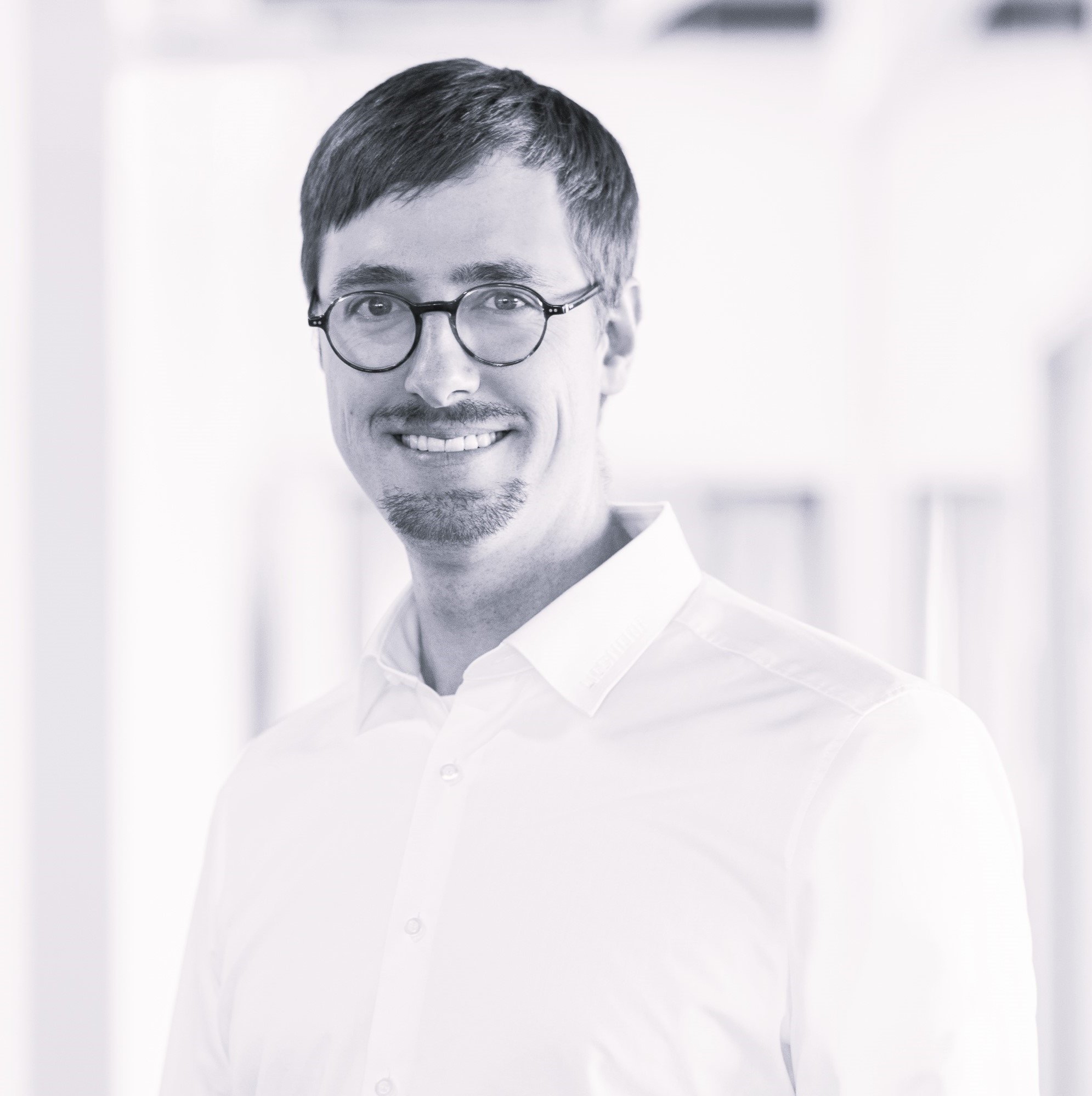}}]{Sebastian Hartwig} is a computer vision researcher and data scientist who completed his master's degree at the Institute of Media Informatics at Ulm University in 2017 before joining the university's Visual Computing Group as a doctoral candidate. His academic work centered on machine scene understanding, monocular depth estimation, synthetic data generation for deep learning, and modeling human perception and preferences. Following his doctoral studies at Ulm University, he transitioned to working as a data scientist at the Liebherr Digital Development Center GmbH.
\end{IEEEbiography}
\vspace{-10mm}
\begin{IEEEbiography}[{\vspace{-0.25in}\includegraphics[width=1in,height=1in]{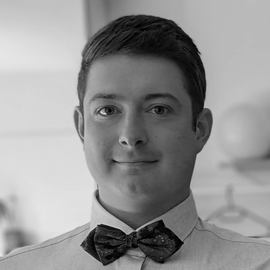}}]{Dominik Engel} is a Research Scientist at King Abdullah University of Science and Technology (KAUST). He received his PhD on "Deep Learning in Volume Rendering" from Ulm University with highest honors in 2024. His research interests in applying Deep Learning to 3D Computer Vision and Computer Graphics.
\end{IEEEbiography}
\vspace{-10mm}
\begin{IEEEbiography}[{\includegraphics[width=1in,height=1.35in]{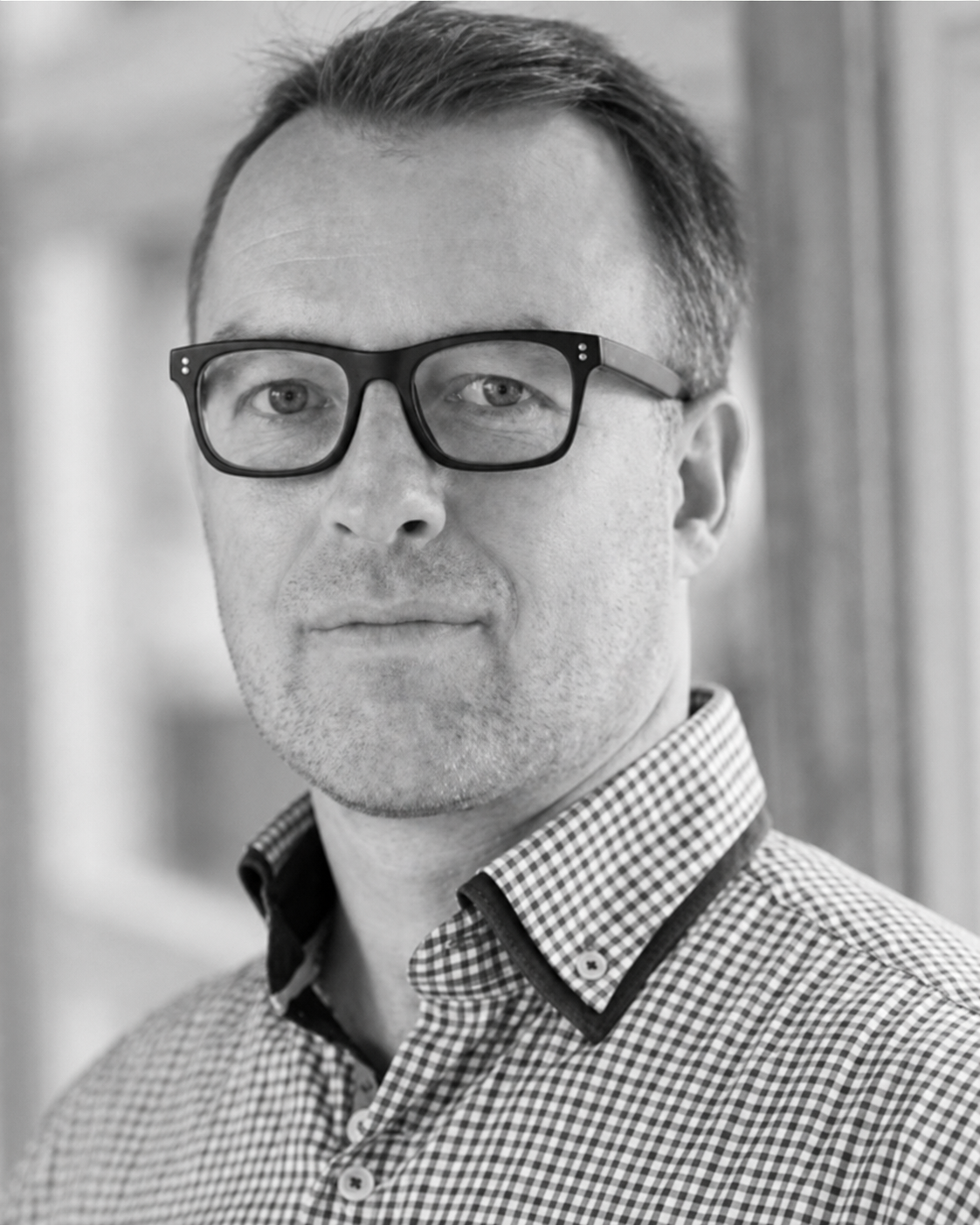}}]{Timo Ropinski} received the PhD degree in computer science from the University of Münster, in 2004, and the habilitation degree from the University of Münster, in 2009. He is a professor with Ulm University, heading the Visual Computing Group. Before moving to Ulm, he was professor in interactive visualization with Linköping University, heading the Scientific Visualization Group. Currently, His serves as chair of the EG VCBM Steering Committee, and as an editorial board member of IEEE TVCG.
\end{IEEEbiography}
\begin{IEEEbiography}[{\includegraphics[width=1in,height=1.25in]{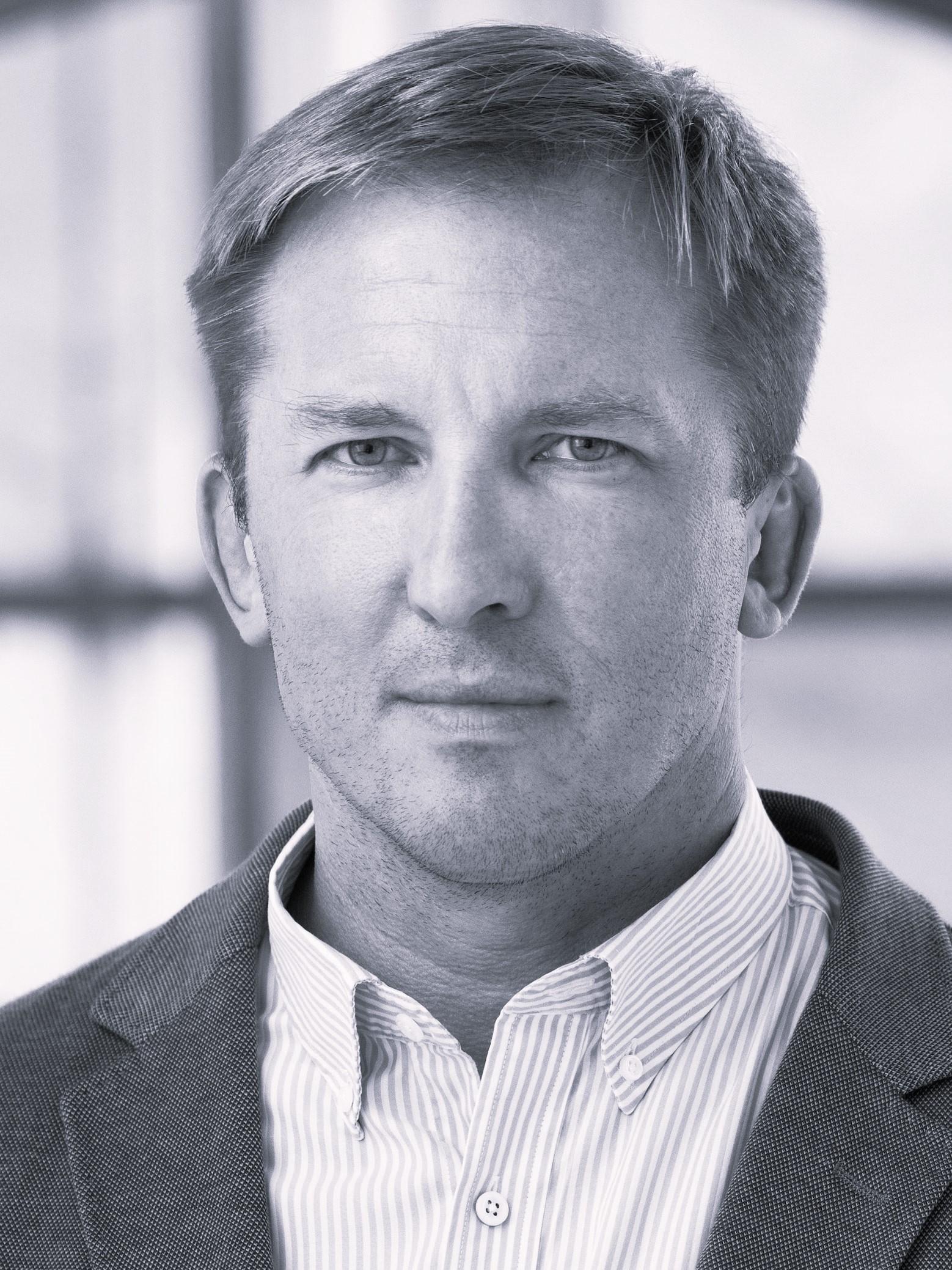}}]{Ivan Viola} is a professor at King Abdullah University of Science and Technology (KAUST), Saudi Arabia. He graduated from TU Wien, Austria. In 2005 he took a postdoc position at the University of Bergen, Norway, where he was gradually promoted to the professor rank. In 2013 he received a WWTF grant to establish a research group at TU Wien. At KAUST, he continues developing new visualization techniques, primarily oriented on data reconstruction, interpretation, representation, modeling, and rendering.
\end{IEEEbiography}
\vspace{0mm}

\end{document}